\documentclass{mscs}

\usepackage{amssymb,amsfonts,latexsym,stmaryrd}
\usepackage{proof}
\usepackage[PostScript=dvips]{diagrams}
\usepackage{my}
\usepackage{graphics}

\title{Sequentiality vs. Concurrency in Games and Logic}
\author[S. Abramsky]{S\ls A\ls M\ls S\ls O\ls N\ns A\ls B\ls R\ls
  A\ls M\ls S\ls K\ls Y\ls\\
Oxford University Computing Laboratory, Wolfson Building, Parks
Road,\addressbreak
Oxford OX1 3QD, U.K.}

\newcommand{\pcomp}{\|}
\newcommand{\leftmerge}{\llfloor}
\newcommand{\linimpl}{\multimap}
\newcommand{\llpar}{\bindnasrepma}
\newcommand{\eqdef}{\; = \;}
\newcommand{\PlusR}{\oplus \mathsf{R}}
\newcommand{\Plusl}{\oplus \mathsf{L}}
\newcommand{\Ten}{\otimes}
\newcommand{\LLPar}{\llpar}
\newcommand{\WITH}{\&}
\newcommand{\Cut}{\mathsf{Cut}}
\newcommand{\Ax}{\mathsf{Id}}
\newcommand{\Foc}{\mathsf{Foc}}
\newcommand{\ShR}{\vdash \downarrow}
\newcommand{\ShL}{\downarrow \vdash}
\newcommand{\TenR}{{\vdash} {\otimes}}
\newcommand{\TenL}{{\otimes} {\vdash}}
\newcommand{\PlusRl}{{\vdash} {\oplus} l}
\newcommand{\PlusRr}{{\vdash} {\oplus} r}
\newcommand{\PlusL}{{\oplus} {\vdash}}
\newcommand{\shneg}[1]{{\downarrow}#1^{\perp}}
\newcommand{\shana}{{\downarrow}({\downarrow} \alpha^{\perp})^{\perp}}
\newcommand{\dnegwith}{\shneg{(\shneg{\alpha} \oplus \shneg{\alpha})}}
\newcommand{\plusform}{\shneg{\alpha} \oplus \shneg{\alpha}}
\newcommand{\Mallf}{\mathrm{MALL}_{\mathsf{foc}}}
\newcommand{\Idproof}{\infer[\Ax]{\girard \alpha^{\perp},  \alpha}{}}
\newcommand{\tensor}{\otimes}
\newcommand{\parc}{\llpar}

\newcommand{\entails}{\girard}
\newcommand{\tuple}[2]{\langle #1 | #2 \rangle}
\newcommand{\inj}{\mathsf{in}}

\newcommand{\eval}[2]{\langle #1 | #2 \rangle}
\newcommand{\up}[1]{#1^{\top}}
\newcommand{\Nat}{\mathbb{N}}
\newcommand{\Bool}{\mathbb{B}}
\newcommand{\MM}{\mathcal{M}}
\newcommand{\LL}{\mathcal{L}}
\newcommand{\GG}{\mathbb{G}}
\newcommand{\Strat}{\mathcal{S}}
\newcommand{\lsem}{[ \! [}
\newcommand{\rsem}{] \! ]}
\newcommand{\intt}{\inj_{\mathtt{tt}}}
\newcommand{\inff}{\inj_{\mathtt{ff}}}

\begin{document}
\maketitle

\begin{abstract}
Connections between the sequentiality/concurrency distinction and the
semantics of proofs are investigated, with particular reference to
games and Linear Logic.
\end{abstract}

\section{Introduction}

We use Games and Logic as a mirror to understand an aspect of the
sequentiality/concurrency distinction. We begin with the simple,
intuitive notion of \emph{polarized games} due to Blass \cite{Bla72,blass}, which prefigured
many of the ideas in Linear Logic \cite{Gir87}, and which can be seen as a polarized
version of ideas familiar from process calculi such as CCS \cite{Mil99} (synchronization trees,
prefixing, summation, the Expansion theorem). We analyze the `shocking'
fact that this very clear and intuitive idea leads to a non-associative
composition; a kind of incompatibility between a purely sequential model
and logic in a classical format. Two ways of addressing this issue have
been found. One is to modify the syntax, by studying a
`hyper-sequentialized' version of sequent calculus, in which the current
focus of attention in the proof is explicitly represented. This is the
approach taken in Girard's Ludics \cite{Gir01}. The other is to broaden the notion of
game to encompass `truly concurrent games'. In such games there is no
longer a global polarization (we can have positions in which both players
can move concurrently), although there are still local polarizations
(each local decision is made by just one of the players). This idea of
concurrent games was used by Abramsky and Melli\`{e}s \cite{AM99} to give a fully
complete model for Linear Logic in its original form (i.e. not
`hyper-sequentialized'), and indeed the correspondence is with
proof-nets, the `parallel syntax for proof theory' in Girard's phrase
\cite{girard:pn}.
In this way, the distinction between sequentiality and concurrency is
reflected at a fundamental level, in the analysis of the `space of
proofs'.

The main aim of the present paper is  to expose some of the
conceptual issues underlying recent technical work on games and logic. The presentation is deliberately
elementary in style, in the hope of making the discussion accessible
both to concurrency theorists, and to those interested in the
semantics of proofs---and of exhibiting a
significant point of contact betwen these two fields.

The analysis of Blass games and the problem of non-associativity of
composition, and the connection of this issue to the
interleaving/true concurrency distinction, were first presented by the
author in a lecture given at the Isaac Newton Institute for the
Mathematical Sciences at Cambridge in 1995, during the programme on
Semantics of Computation.

For the reader's convenience, some material on 
Linear Logic (specifically, the sequent calculus presentation of
propositional Multiplicative-Additive Linear Logic (MALL)) is
recalled in an appendix.

\paragraph{Acknowledgement}
The comments made by the journal referees suggested a number of
improvements to the presentation.
\section{Blass Games as Polarized Processes}
In this section we will view games as \emph{polarized processes}. More
precisely, we will develop a correspondence between certain 2-person games of
perfect information and polarized deterministic synchronization trees.

\subsection{Review of notions from process theory}
We can describe (well-founded) deterministic synchronization trees
inductively, as given by expressions of the form
\[ \sum_{i \in I} a_i . P_i \quad \quad (i \neq j \; \Rightarrow 
\; a_i \neq a_j ) \]
i.e. as the least set closed under the operation of \emph{disjointly
  guarded summation} \cite{Mil89}. It is understood that the
summation, as in CCS, is associative and commutative (idempotence does 
not arise because of the disjointness condition). The basic case of
the inductive definition is given by the empty summation, written $0$
(the `NIL' process of CCS). There are labelled transitions
\[ \sum_{i \in I} a_i . P_i \stackrel{a_i}{\longrightarrow} P_i \]
for each $i \in I$, giving the arcs from the root of the
synchronization tree to its immediate sub-trees. See \cite{WN95} for
useful background on synchronization trees and related models.

We could accomodate infinite branches in such trees by using a
coinductive rather than an inductive definition. This issue is not
important for our purposes here, for which it will be quite sufficient to
consider only \emph{finite} trees.

We can define interleaving (non-communicating) parallel composition on 
the synchronization trees thus: if $P =  \sum_{i \in I} a_i .
P_i$, and $Q  =   \sum_{j \in J} b_j .
Q_j$, then
\[ P \pcomp Q \; = \; \sum_{i \in I} a_{i}. (P_i \pcomp Q) + \sum_{j
  \in J} b_{j}. (P \pcomp Q_{j}) . \]
(In order to preserve the disjointness property in the summation, we
require that the \emph{sorts} of $P$ and $Q$ (i.e. the sets of labels
appearing anywhere in the synchronization trees for $P$ and $Q$ respectively)
are disjoint \cite{Mil89}---this will
be tacitly assumed in the sequel.)

This is the Expansion Theorem \cite{Mil89} in an appropriate version. Note
that it can be taken as an (inductive or coinductive)
\emph{definition} of parallel composition as an operation on
synchronization trees. It shows how to \emph{eliminate} parallel
composition in favour of purely sequential constructs. As such, it
expresses the essence of the interleaving view of concurrency.
Note that, in Milner's classification of the operations of process
algebra \cite{Mil89}, guarded summation is built from the \emph{dynamic}
operations, while parallel composition is the key \emph{static}
operation. So the interleaving view gives primacy to the dynamic
operations.

It will also be useful to recall the \emph{left merge} operation,
which is used  extensively in Process Algebra \cite{BW90}. This has the
defining equation
\[ P \leftmerge Q \; = \; \sum_{i \in I} a_{i}. (P_i \pcomp Q) \]
where $P = \sum_{i \in I} a_i .
P_i$, and we can then express the Expansion Theorem by
\[ P \pcomp Q \; = \; P \leftmerge Q + Q \leftmerge P. \]
This `biassed scheduling' will also turn out to have its counterpart 
in Games and Linear Logic.
\subsection{Game Trees}
We now consider game trees for 2-person games of perfect
information. We will represent such games by \emph{polarized}
deterministic synchronization trees, i.e. by trees with one `bit' of
information added at each node to say which of the two player's turn
it is to play at the game position corresponding to that node. We
follow a standard practice of referring to the two players as P
(`Proponent') and O (`Opponent'). In process terms, we can think
of P as representing the System currently under consideration, and O
as its Environment (see \cite{gfc} for a more detailed account of the
correspondence between notions of game theory and process theory).

Formally, we take games to be inductively (or, if preferred,
coinductively) defined to be either a \emph{product of games} or a
\emph{coproduct of games}
\[ \prod_{i \in I} G_i \quad \mathrm{or} \quad \coprod_{i \in I} G_i
\]
where
\[ \prod_{i \in I} G_i \eqdef (\sum_{i \in I} i . G_i , O) \]
\[ \coprod_{i \in I} G_i \eqdef (\sum_{i \in I} i . G_i , P) \]
(Since the identities of the `actions' in the synchronization trees
are irrelevant---they just label the possible moves in each position
of the game---we use the indices in the summations themselves as
the action names.)

There are now two versions of the empty game tree: the empty product
$(0, O)$, and the empty coproduct $(0, P)$.

Clearly, we can define a formal `game negation' which inverts the
polarization at each node of the tree:
\[ (\sum_{i \in I} i . G_i , \pi )^{\bot} \eqdef (\sum_{i \in I}
i . G_i^{\bot} , \bar{\pi} ) \]
where $\bar{\mathrm{P}} = \mathrm{O}$, $\bar{\mathrm{O}} =
\mathrm{P}$. 
This immediately yields the equations

\[ \begin{array}{lcl}
(\prod_{i \in I} G_i )^{\bot} & \eqdef  & \coprod_{i \in I} G_i^{\bot} \\
(\coprod_{i \in I} G_i )^{\bot} & \eqdef &  \prod_{i \in I} G_i^{\bot} \\
G^{\bot \bot}  & \eqdef &  G . 
\end{array} \]

In Linear Logic, the product and coproduct are the \emph{additive}
level of connectives, with the product written as $\&$ (With) and the
coproduct as $\oplus$ (Plus). We see that in terms of the game trees,
these connectives are accounted for by the \emph{dynamic
  operations}. What about the static operations, in particular
parallel composition? This has two polarized versions, corresponding
to the \emph{multiplicative} connectives $\otimes$ (Tensor) and
$\llpar$ (Par) of Linear Logic.

\subsection{Tensor}

To define each of these binary connectives, we must consider four cases,
corresponding to the possible polarizations of their arguments.

For the Tensor, three of the four cases are forced by the requirement
that Tensor should distribute over coproduct. This can be taken as a desirable
logical distributivity; more mathematically, to support a good notion
of implication, Tensor with either argument fixed should be a left
adjoint, giving rise e.g. to the adjunction
\[ \infer={A \; \longrightarrow \; B \linimpl C}{A \otimes B \; 
 \longrightarrow \; C} \]
and hence must preserve all colimits. In game terms, thinking 
of Tensor as a form of conjunction, the onus is on P to defend both
conjuncts, and hence P should move whenever it is his term to move in
either of the two ``sub-games''. Thus we obtain the following
`equations' for the tensor:
\[ \begin{array}{lcl}
\coprod_{i \in I} G_i \otimes H & \eqdef & \coprod_{i \in I} (G_i \otimes H) 
\\
G \otimes \coprod_{j \in J} H_j & \eqdef & \coprod_{j \in J} (G \otimes H_{j}) 
\end{array} \]
As it stands, these equations are ambiguous, since they might both
apply.
We therefore only apply them in the case where the term not appearing
as a coproduct (i.e. $H$ in the first equation and $G$ in the second)
is labelled by O at the root (`has negative polarity').
To cover the final case, where both arguments to the tensor are
coproducts, we write
\[ \coprod_{i \in I} G_i  \otimes \coprod_{j \in J} H_j \eqdef
\coprod_{(i, j)
  \in I \times J} (G_i \otimes H_{j}) 
\]
Here the right-hand side introduces the index set $I \times J$,
indicating that the initial move consists of P \emph{simultaneously} moving
in both sub-games. This is essentially the convention followed by
Blass \cite{blass}. Alternatively, we might think of the moves being
performed in either order (an interleaving of the moves $i$ and
$j$). These representational issues are not crucial here. What is
important is the specification that P must move in the tensor whenever 
it is his turn to move in either sub-game.

In process terms we can think of these cases as `biassed
scheduling', of the form of the \emph{left merge} operator studied in 
Process Algebra, with the bias  determined by the polarity.

The final case to be considered is when the polarities of both the
arguments to the Tensor are O: and here we simply apply the exact
analogue of the Expansion Theorem for synchronization trees. If $G =
\prod_{i \in I} G_i$, and  $H =
\prod_{j \in J} H_j$, then:
\[ G \otimes H \eqdef \prod_{i \in I} (G_i \otimes H) \; \sqcap \; \prod_{j \in
  J} (G \otimes H_{j}) . \]
(Here we write $\sqcap$ for binary product).
In game terms, this says that O has the freedom to move in either
sub-game, which may then restore the situation to one in which P will
be forced to move; this is analogous (and indeed, formally related) to
the `switching conditions' introduced in \cite{gfc}.

It is worth pointing out that this case, unlike the previous ones, is
not logically (or mathematically) \emph{forced} by the wish to
obtain a distributivity property or adjunction.
Rather, it is being enforced by the commitment in the setting we are
currently considering to a strictly sequential schedule of events,
where it is always exactly one player's turn to move. Indeed, when we
move to our concurrent form of games, we shall allow a wider class of
behaviours.

Another point worth making is that the third equation (both arguments 
coproducts) can never be reached from one of the other cases. Whenever
the polarities at the roots of the two sub-games being considered are
any of
(O,P), (P,O) or (O,O), then any of the initial actions which may be
performed lead to a residual pair of games whose root polarities are again one
of these three. For a discussion and proof of this, see \cite{Abr97a}.

It may also be useful to give a `polarity table' for the tensor, to show 
how the root of a tensor game is labelled as a function of the labels
of its subgames.
\begin{center}
\begin{tabular}{c|cc}
\hline \hline
$G$ & $H$ & $G \otimes H$ \\ \hline
P & P & P \\ 
P & O & P \\ 
O & P & P \\ 
O & O & O \\ 
\hline \hline
\end{tabular}
\end{center}

This can be compared to the truth-table for conjunction (reading O as
true and P as false), which
appears to have been part of Blass's original motivation for his
definition of the tensor.

A similar table appears in \cite{Lam95} (with references to earlier
work by Danos, Regnier, Malacaria and Bellin), with the salient difference
that the case for (O, O) is not defined there. This reflects the fact
that in that paper, it is really Intuitionistic Linear Logic which is
the object of study, and Classical Linear Logic is studied indirectly
via a form of double-negation translation.
This can be seen as a special case of the restrictions imposed by the focussing version of Linear
Logic, to be discussed in Section~6.

\subsection{Par and Linear Implication}
Given that we wish to have the De Morgan Duality
\[ G \llpar H \eqdef (G^{\bot} \otimes H^{\bot})^{\bot} \]
and the equivalence
\[ G \linimpl H \eqdef G^{\bot} \llpar H \]
then the definitions of the other multiplicative connectives are forced 
by that of the tensor:
\[ \begin{array}{lcl}
\prod_{i \in I} G_i \llpar H & \eqdef & \prod_{i \in I} (G_i \llpar H) 
\\
G \llpar \prod_{j \in J} H_j & \eqdef & \prod_{j \in J} (G \llpar H_{j}) 
\\
\prod_{i \in I} G_i  \llpar \prod_{j \in J} H_j & \eqdef & \prod_{(i,
  j) \in I \times J} (G_i \llpar H_{j}) 
\end{array}
\]
and if $G =
\coprod_{i \in I} G_i$, and  $H =
\coprod_{j \in J} H_j$, then:
\[ G \llpar H \eqdef \coprod_{i \in I} (G_i \llpar H) \; \sqcup \; \coprod_{j \in
  J} (G \llpar H_{j}) . \]
(Here we write $\sqcup$ for binary coproduct).
The reader can similarly write down the definition for the Linear
Implication.

This notion of game tree, and the interpretations of the
multiplicative and additive  connectives of Linear Logic, correspond
exactly to the game semantics of Andreas Blass \cite{blass} (in the
`relaxed form' in his terminology). Note that the additives provide
the operations by which all such game trees can be constructed, and
the multiplicatives are `eliminated' by the polarized versions of
the Expansion Theorem given above. Hence the additives have primacy in 
this form of game semantics, in exactly the same sense that the dynamic
operations have primacy in the interleaving view of process algebra.

These definitions are very clear and natural; indeed, they seem
forced, given that the interleaving point of view is being taken---as
it is, implicitly at least, in the great bulk of the tradition in Game Theory, which
concerns game trees in which play proceeds in a purely sequential
fashion.

\subsection{The second level: strategies}
As we have seen, games add one crucial bit of structure to processes,
namely the information about whose turn it is to play at each
position. This opens the way to allowing a great deal of additional
structure to be articulated explicitly. The usual process models are
rather amorphous; they allow a great variety of behaviours to be
expressed easily, but the structural characteristics of various
classes of behaviour are hard to capture.

For example, processes are used to model both specifications and
implementations: both types, and the inhabitants of those types. In
the world of games, one has the basic distinction between games and
strategies. If games are represented as trees, then one can naturally
represent strategies for P or O for a game $G$ as certain kinds of
subtree of $G$. From the point of view of process theory, both games
and strategies are processes, but we can use the distinction offered
by the game setting to build a more structured theory.

We exemplify this by defining O- and P-strategies for Blass games. For 
simplicity we shall consider only deterministic, total strategies for
well-founded games,
which have a defined response in every position which can arise. For
$\pi \in \{ P, O \}$ we define $S^{\pi}(G)$, the strategies for $\pi$
on the game $G$, inductively by:
\[ \begin{array}{lcl}
S^{\pi}(\sum_{i \in I} i . G_i , \pi ) & = & \{ \bar{i} .
\sigma_i \mid i \in I \wedge \sigma_i \in S^{\pi}(G_{i}) \} \quad (I
\neq \varnothing ) \\
S^{\pi}(0, \pi ) & = & \{ 0 \} \\
S^{\pi}(\sum_{i \in I} i . G_i , \bar{\pi} ) & = & \{ \sum_{i
  \in I} i .
\sigma_i \mid \forall i \in I .   \, \sigma_i \in S^{\pi}(G_{i}) \}
. 
\end{array} \]
Thus a strategy for player $\pi$ on a game $G$ is a
sub-synchronization tree of $G$ which makes a deterministic choice of
move at each position where $\pi$ is to play in $G$, and has a
response for every possible move by $\bar{\pi}$ where it is
$\bar{\pi}$'s turn. Note that, following CCS \cite{Mil89}, we
represent the player $\pi$'s own moves by \emph{output actions}
$\bar{i}$, and the moves that $\pi$ `observes' $\bar{\pi}$ to make by
\emph{input actions} $i$.

Given $\sigma \in S^{P}(G)$,  $\tau \in S^{O}(G)$, we can define the
result of playing $\sigma$ off against $\tau$ to be
\[ (\sigma \mid \tau )\setminus S \]
using the CCS parallel composition and restriction operators
\cite{Mil89},
where $S$ contains all the actions appearing in $G$. This will result in 
a uniquely determined silent computation, which for well-founded $G$
will end in an empty node $(0, \pi )$, where player $\pi$ is to play
but no move is possible. We decree that player $\pi$ loses that
play; and can then proceed to define the notion of winning strategy in the
usual fashion. For non-well-founded game trees, infinite plays are
possible, and we must add information to say who wins these.
For more on this, see \cite{Abr97a}, and for further development of the
connection between processes, games and Linear Logic, see \cite{Abr99}.
We omit a detailed discussion of these ideas, which are not needed for our
purposes in this paper.

\section{Composition}
We now turn to the crucial issue of \emph{composition} of
strategies. It seems that, with our simple but highly suggestive
refinement from processes into Blass games, we have all the necessary 
ingredients to make a full-blown model of Multiplicative-Additive
Linear Logic (MALL), at the level, not just of formulas, but of
\emph{proofs}. More precisely, we can see a clear idea for how to
define a category $\mathcal{G}$ based on games and strategies which
can serve as a model for MALL. The objects of $\mathcal{G}$ will be
Blass games; a morphism from $A$ to $B$ will be a strategy $\sigma \in 
S^{P}(A \linimpl B) = S^{P}(A^{\perp} \llpar B)$. We have defined interpretations of tensor
product ($A \otimes B$), par ($A \llpar B$), negation ($A^{\bot}$), products ($\prod_{i
  \in I} A_i$) and coproducts ($\coprod_{i \in I} A_i$) in this
category. This is all so natural that one might take it for granted
that the details must all work out.

However, there is a snag, of a rather fundamental kind. Moreover, this 
problem arises precisely from the polarization structure we have imposed 
on the Linear connectives, and hence from the interleaving view of
concurrency on which, implicitly, it is based.

The problem concerns \emph{associativity of composition}. It was known 
to Andreas Blass \cite{blasspc}, although not mentioned in his
papers on Game Semantics. It was noticed independently and discussed in 
\cite{gfc}.

\noindent To analyze associativity, we should consider strategies
\[ \sigma : A \rightarrow B \qquad \tau : B \rightarrow C \qquad \theta
: C \rightarrow D \]
which we can view as
\[ \sigma : A^{\bot} \llpar B \qquad \tau : B^{\bot} \llpar C \qquad \theta
: C^{\bot} \llpar D . \]
Each of the games $A$, $B$, $C$, $D$ can have either P or O polarity
at the root, yielding 16 possibilities. Of these, \emph{just one is
  problematic}, namely when:
\begin{itemize}
\item $A$, $C$ have polarity O
\item $B$, $D$ have polarity P.
\end{itemize}
In this case, $A^{\bot}$ has polarity P, so in the game $ A^{\bot}
\llpar B$ we are in the case of the Expansion Theorem for Par, and P
has the choice of playing in either $A^{\bot}$ or $B$. Suppose that
$\sigma$ in fact chooses to play in $A^{\bot}$. Similarly,  $C^{\bot}$ has polarity P, so in the game $ C^{\bot}
\llpar D$,  P
has the choice of playing in either $C^{\bot}$ or $D$. Suppose that
$\theta$ in fact chooses to play in $D$. Now in $B^{\bot} \llpar C$, O 
must play first. Suppose that we first form the composition $\sigma ;
\tau : A^{\bot} \llpar C$. Then again O is to play first in this
game. If we now form the composition 
\[ (\sigma ; \tau ) ; \theta : A^{\bot} \llpar D , \]
then since P is to play first, the only possibility is that the first
move is that of $\theta$, which is made in $D$. Now reasoning in an
entirely similarly fashion, if we first form the composition $\tau ;
\theta$, this must first wait for O to move, while $\sigma ; (\tau ;
\theta )$ must begin with a move by P, which must then be made by
$\sigma$, and hence must be made in $A^{\bot}$.
We therefore conclude that:
\begin{itemize}
\item $(\sigma ; \tau ) ; \theta$ plays initially in $D$ 
\item  $\sigma ; (\tau ;
\theta )$ plays initially in $A^{\bot}$
\end{itemize}
and hence $(\sigma ; \tau ) ; \theta \neq \sigma ; (\tau ;
\theta )$.

This argument is very robust, and assumes very little about the
specifics of how strategies are defined. It is based on the way that
polarization is enforced in the definition of the multiplicative
connectives in Blass games. As we have seen, that enforcement of
polarization is rather directly related to the interleaving
interpretation of concurrency.

What of the other 15 possible polarizations of $A$, $B$, $C$, $D$? We
can see fairly easily that none of them give rise to an analogous
problem. Indeed, in 12 of these cases, O is to move first in $A^{\bot}
\llpar D$; while in the three remaining cases, one of the three
strategies $\sigma$, $\tau$ or $\theta$ must unambiguously move first
as forced by the polarization, whichever way the composition is
bracketed. We list the possibilities explicitly in Table~1.
\begin{table}
\begin{tabular}{|l|l|l|l|l|c|} \hline
 & A & B & C & D & initial move by \\ \hline
0 & O & O & O & O & O \\ \hline
1 & O & O & O & P & $\theta$ \\ \hline
2 & O & O & P & O & O \\ \hline
3 & O & O & P & P & $\tau$ \\ \hline
4 & O & P & O & O & O \\ \hline
5 & O & P & O & P & ? \\ \hline
6 & O & P & P & O & O \\  \hline
7 & O & P & P & P & $\sigma$ \\ \hline
8 & P & O & O & O & O \\ \hline
9 & P & O & O & P & O \\ \hline
10 & P & O & P & O & O \\ \hline
11 & P & O & P & P & O \\ \hline
12 & P & P & O & O & O \\ \hline
13 & P & P & O & P & O \\ \hline
14 & P & P & P & O & O \\ \hline
15 & P & P & P & P & O \\ \hline
\end{tabular}
\caption{How polarities determine the initial move}
\end{table}

\section{Syntactic Perspective}
\label{syntpers}
To show how easily the problem we described in the previous section
can arise in MALL, we shall give a simple example of a proof in MALL
whose interpretation in Blass games gives rise to exactly this situation.

Firstly, let $\alpha$ be a propositional atom. We define $A$, $B$, $C$ 
and $D$ to be the following formulas:
\[ A = C = \alpha \& \alpha , \qquad B = D =  \alpha \oplus \alpha
. \]
Note that these formulas have exactly the polarities of the
problematic case discussed in the previous section (remembering that
$\&$ is product and $\oplus$ is coproduct).
We define three proofs.

\[ \Pi_1 : \qquad
\infer[\PlusR]{\girard \alpha^{\perp} \oplus \alpha^{\perp}, \alpha
  \oplus \alpha}{
\infer[\PlusR]{\girard \alpha^{\perp},  \alpha
  \oplus \alpha}{
\Idproof } }  \]

\[ \Pi_2 : \qquad
\infer[\WITH]{\girard \alpha^{\perp} \& \alpha^{\perp}, \alpha
  \& \alpha}
{\infer[\WITH]{\girard \alpha^{\perp} \& \alpha^{\perp}, \alpha}
  {\Idproof  & \Idproof}
 & 
\infer[\WITH]{\girard \alpha^{\perp} \& \alpha^{\perp}, \alpha}
  {\Idproof  & \Idproof}
}
 \]

\[ \Pi_3 : \qquad
\infer[\PlusR]{\girard \alpha^{\perp} \oplus \alpha^{\perp}, \alpha
  \oplus \alpha}{
\infer[\PlusR]{\girard \alpha^{\perp} \oplus \alpha^{\perp},  \alpha}
{\Idproof} }  \]

The interpretation of these proofs as strategies $\sigma$, $\tau$,
$\theta$ gives rise to exactly the problematic situation described in
the previous section. The relationship between the form of the sequent 
calculus proofs and the temporal properties of strategies (where do we 
move first?) is that 
\begin{center}
\emph{the \textbf{last rule} of a proof corresponds to 
  the \textbf{first move} in the associated strategy}.
\end{center} 
Thus  the first proof
above, $\Pi_1$, indeed `moves' first in $A^{\perp}$, while the third
proof $\Pi_3$ moves first in $D$. On the other hand, rules introducing 
connectives of negative polarity, such as $\&$, correspond to O-moves; 
so we see that in the second proof, $\Pi_2$, O is to move first.

Now we can consider the two ways of composing these proofs using cuts:
\[ \infer[\Cut]{\girard A^{\perp}, D}
 { \infer[\Cut]{\girard A^{\perp}, C}
              {\infer*{\girard A^{\perp}, B}{\Pi_{1}}
              & \infer*{\girard B^{\perp}, C}{\Pi_{2}}
              }
 & \infer*{\girard C^{\perp}, D}{\Pi_{3}}
 } \]

\[ \infer[\Cut]{\girard A^{\perp}, D}
 { \infer*{\girard A^{\perp}, B}{\Pi_{1}}
 & 
 \infer[\Cut]{\girard B^{\perp}, D}
              {\infer*{\girard B^{\perp}, C}{\Pi_{2}}
              & \infer*{\girard C^{\perp}, D}{\Pi_{3}}
              }
 } \]
corresponding to $(\sigma ; \tau );\theta$ and $\sigma ; (\tau ;
\theta )$ respectively.
Performing cut-elimination on these two proofs leads to $\Pi_{3}$ and
$\Pi_{1}$ repectively. These two proofs can be considered equal
\emph{up to permutation of rules}. The
`rigid' sequential interpretation of proofs as strategies does not
satisfy this equality, and hence distinguishes the two
proofs.

The analogy we wish to make is again with concurrency theory. The two
proofs differ in the order in which they introduce the two occurences
of $\oplus$ in the sequent $\girard \alpha^{\perp} \oplus \alpha^{\perp}, \alpha
  \oplus \alpha$. Computationally, each of these introduction rules
  amounts to specifying one bit of information (are we in the left
  disjunct or the right disjunct?), so that we can see the r\^{o}le of these two
  introduction rules in the proof as like performing two `actions', say:
\begin{itemize}
\item $a$: set the boolean variable $B_1$
\item $b$: set the boolean variable $B_2$.
\end{itemize}
From the sequential, interleaving point of view, we must
distinguish between the computation histories $ab$ and $ba$, taking
account of the order in which the `events' $a$ and $b$ occur. In a
true concurrency model, if the events $a$ and $b$ are causally
independent of each other, we can identify these two histories \cite{WN95}.

The purpose of proof nets \cite{Gir87,girard:pn} and Geometry of Interaction \cite{Gi89,Gi90,Gi95,DR93,DR95}
    was to find a more intrinsic representation of proofs in which the 
    spurious ordering of rules imposed by sequent calculus was
    factored out. This is directly akin to the issue of representing true
    concurrency; and we can indeed see proof nets as a `parallel
    syntax for proof theory' \cite{girard:pn}.

We conclude this section by remarking that  the syntactic example we
have discussed above can be linked directly to the general analysis of 
the associativity problem for Blass games given in the previous
section. One simply has to spell out the interpretation of the three
proofs $\Pi_1$, $\Pi_2$, $\Pi_3$ as strategies $\sigma_1$, $\sigma_2$, 
$\sigma_3$ for the corresponding
Blass games, and observe that 
\[ \sigma_1 ; (\sigma_2 ; \sigma_3 ) \neq (\sigma_1 ; \sigma_2 ) ;
\sigma_3 . \]
Exactly this is done in \cite{gfc}. We prefer not to repeat it here
since we have not given a formal definition of composition of
strategies for Blass games.
\section{Discussion}
It is useful to compare the issue we have raised in connection with
Classical Linear Logic with a familiar issue in the proof theory of
Classical Logic. There is a well known incompatibility between
unconstrained Classical Logic proof theory and \emph{confluence}, as
shown very clearly and simply by a well-known example due to Yves
Lafont \cite{GLT}. In our view, the example we have discussed shows
equally clearly the incompatibility between unconstrained Classical
Linear Logic proof theory and a purely sequential, interleaving view
of the semantics of proofs. 

There seem to be two approaches to the computational interpretation of 
proofs in Classical Logic.

\begin{enumerate}
\item `Tame' the syntax, by restricting the permitted reductions, or
  by adding information to resolve the non-deterministic choices, and
  hence restoring confluence. This approach is typified by systems
  such as Girard's LC \cite{Gir91}, the systems LKT and LKQ of Danos, Joinet and Schellinx
  \cite{DJS}, and the $\lambda \mu$-calculus of Parigot \cite{Par92}.

\item Study the full, `untamed' non-confluent calculus, and find the
  computational structure which is there. This is typified by work
  such as \cite{BB96,urbanphd,Lai01}.
\end{enumerate}

In a similar fashion, there seem to be two possibilities for studying
Linear Logic proofs.
\begin{enumerate}
\item Taming the syntax, in such a way as to control the scheduling,
  and to avoid the `bad' combination of polarities which led to the
  problem with composition.
Most game semantics applies to the \emph{Intuitionistic} version of
Linear Logic---indeed much of it applies only to the negative fragment 
\cite{Abr97a}. In game terms, this means that one only considers games in which 
Opponent starts, and in which the players strictly alternate. Under
these constraints, no problems with composition arise. What is lost is 
any direct access to the natural game interpretation of classical
(Linear) negation, by interchange of r\^{o}les between the players.
However, it should be noted that these kind of games are quite sufficient to yield good models 
of the $\lambda \mu$-calculus, as shown e.g. in \cite{Ong96,Lai99}.

But this is not the only option. One can go beyond the scope of
Intuitionistic Linear Logic, and study a more expressive, yet still
`tamed' system of Linear Logic. This is Linear Logic with focussing,
as introduced by Andreoli \cite{AP91}, and used in the current work by
  Girard on Ludics \cite{Gir01}. We shall study this approach in the next
  section.

\item Alternatively, we can extend the semantics to embrace true
  concurrency, and hence overcome the problems with composition while
  still interpreting the full, unconstrained original syntax of Linear
  Logic. This is what is done in the concurrent games model, which we
  shall discuss in Section~\ref{cgsec}.
\end{enumerate}

\section{The sequential way out: Focussing}
We shall consider the focussing version of MALL used in \cite{Gir01}, and
  based on the system introduced in \cite{AP91}. We shall refer to this system as
    $\Mallf$. We shall only use the propositional part of the
    system, which we recapitulate here for the reader's convenience.

Firstly, the formulas of the system are \emph{positive formulas} $P$,
$Q$ \ldots, built from propositional atoms 
$\alpha$, $\beta$, \ldots by the grammar
\[ P \;\; ::= \;\; \alpha \mid \shneg{P} \mid P \oplus P \mid P
\otimes P . \]
Here $\shneg{P}$ is the `shifted negation' of $P$, in which the shift
is used to restore positive polarity. One can think of this as a kind
of double negation interpretation: in game/process terms, it is
\emph{polarized prefixing}, in which the polarity is reversed at each
node of the game for $P$ (as in Linear negation), but then positive
polarity is restored by prefixing with a `dummy' P-move.

A sequent in $\Mallf$ is an expression $\Gamma \vdash \Delta ;
\Sigma$, where $\Gamma$, $\Delta$, $\Sigma$ are finite multisets of
formulas, with the following \emph{stoup constraint}:
\begin{center}
\textit{If the stoup $\Sigma$ is non-empty, then it contains exactly one
formula; moreover, in this case $\Gamma$ consists only of
propositional atoms.}
\end{center}
Negative propositions are handled implicitly through the left hand
sides of sequents.

\noindent The rules of $\Mallf$ are as follows.
\\
\noindent \textbf{Cut}
\[ \infer[;\Cut]{\Gamma , \Gamma' \vdash \Delta , \Delta' ;}
{ \Gamma \vdash \Delta ; P & \Gamma' , P \vdash \Delta' ;}
\qquad
\infer[\Cut ;]{\Gamma , \Gamma' \vdash \Delta , \Delta' ;}
{ \Gamma \vdash \Delta , P; & \Gamma' , P \vdash \Delta' ;}
\]
\textbf{Identity}
\[ \infer[\Ax]{\alpha \vdash ; \alpha}{} \]
\textbf{Focussing}
\[ \infer[\Foc]{\Gamma \vdash \Delta , P;}{\Gamma \vdash \Delta ; P}
\]
\textbf{Shift}
\[ \infer[\ShR]{\Gamma \vdash \Delta ; \shneg{P}}{\Gamma , P \vdash
  \Delta ;}
\qquad
\infer[\ShL]{\Gamma , \shneg{P} \vdash \Delta ;}{\Gamma \vdash
  \Delta , P;}
\]
\textbf{Tensor}
\[ \infer[\TenR]{\Gamma , \Gamma' \vdash \Delta , \Delta' ; P \otimes
  Q}
{\Gamma \vdash \Delta ; P & \Gamma' \vdash \Delta' ; Q}
\qquad
\infer[\TenL]{\Gamma, P \otimes Q \vdash \Delta ;}{\Gamma , P, Q
  \vdash \Delta ;}
\]
\textbf{Plus}
\[ \infer[\PlusRl]{\Gamma \vdash \Delta ; P \oplus Q}{\Gamma \vdash
  \Delta ; P}
\qquad
\infer[\PlusRr]{\Gamma \vdash \Delta ; P \oplus Q}{\Gamma \vdash
  \Delta ; Q}
\]
\[ \infer[\PlusL]{\Gamma , P \oplus Q \vdash \Delta ;}
{\Gamma , P \vdash \Delta ; & \Gamma , Q \vdash \Delta ;}
\]
Note that the stoup represents the ``focus of attention'' for positive
rules (i.e. for P-moves). The stoup constraint enforces an explicit
scheduling, in the sense that, if the stoup is non-empty, then the
only rules which can be applied (backwards, i.e. in the sense of proof 
search) apply to the stoup formula. On the other hand, if the stoup is 
empty, we must apply negative rules to formulas on the left-hand side
of the sequent. So  along each branch of a partial proof 
tree developed in proof search we observe an alternation between a
sequence of P-moves (corresponding to applying rules to the stoup
formula) and O-moves (when the stoup is empty and we apply negative
rules). Positive rules are applied until we hit a shift in the stoup
formula, at which point the stoup is emptied, and the formula
transferred (with reversed polarity) to the left-hand side of the
sequent, so that it is now ``O's turn''. In the opposite direction, O
moves as long as it can, and then  transfers control back to P when all the
formulas on the left-hand side are atoms. P can then refill the stoup
using the focussing rule, and proceed. Thus we see that in this
system, the scheduling of whose turn it is to move next is completely
deterministic, and all actions can be performed in a purely sequential
fashion.

\noindent We now ask: to what extent can the problematic situation in standard
MALL involving the proofs $\Pi_1$, $\Pi_2$, $\Pi_3$ which we discussed in
section~\ref{syntpers} be replicated in $\Mallf$?

If we begin with $\Pi_1$, we can certainly find a reasonable analogue
$\Pi'_{1}$:
\[ \infer[\Foc]{\vdash \shneg{\alpha} \oplus \alpha , \alpha \oplus
  \alpha ;}
{ \infer[\PlusRr]{\vdash \alpha \oplus \alpha ;  \shneg{\alpha} \oplus
    \alpha}
{ \infer[\ShR]{\vdash \alpha \oplus \alpha ;  \shneg{\alpha}}
{ \infer[\Foc]{\alpha \vdash \alpha \oplus \alpha ;}
{ \infer[\PlusRr]{\alpha \vdash ; \alpha \oplus \alpha}
{ \infer[\Ax]{\alpha \vdash ; \alpha}{}
} } } } }
\]
This has the essential feature that the last positive rule is applied
to the left hand formula in the final sequent, which will not participate in the Cut with (the
analogue of) $\Pi_2$.

Now we must find an analogue of $\Pi_2$. This is harder work, and we
must use the `double negation' style representation of $\alpha \&
\alpha$ as $\shneg{(\shneg{\alpha} \oplus \shneg{\alpha})}$. We then
obtain $\Pi'_{2}$:
\[ \infer[\PlusL]{\alpha \oplus \alpha \vdash \dnegwith ;}
{ \infer[\Foc]{\alpha \vdash \dnegwith ;}
 {\infer[\ShR]{\alpha \vdash ; \dnegwith}
  { \infer[\PlusL]{\alpha , \shneg{\alpha} \oplus \shneg{\alpha} \vdash ;}
    { \infer[\ShL]{\alpha , \shneg{\alpha}  \vdash ;}
      { \infer[\Foc]{\alpha  \vdash \alpha ;}
        {  \infer[\Ax]{\alpha  \vdash ; \alpha}{}
        }
      }
    & 
      \infer[\ShL]{\alpha , \shneg{\alpha}  \vdash ;}
      { \infer[\Foc]{\alpha  \vdash \alpha ;}
        {  \infer[\Ax]{\alpha  \vdash ; \alpha}{}
        }
      }
    }
  }
 }
&
  \infer[\Foc]{\alpha \vdash \dnegwith ;}
  {\infer[\ShR]{\alpha \vdash ; \dnegwith}
  { \infer[\PlusL]{\alpha , \shneg{\alpha} \oplus \shneg{\alpha} \vdash ;}
    { \infer[\ShL]{\alpha , \shneg{\alpha}  \vdash ;}
      { \infer[\Foc]{\alpha  \vdash \alpha ;}
        {  \infer[\Ax]{\alpha  \vdash ; \alpha}{}
        }
      }
    & 
      \infer[\ShL]{\alpha , \shneg{\alpha}  \vdash ;}
      { \infer[\Foc]{\alpha  \vdash \alpha ;}
        {  \infer[\Ax]{\alpha  \vdash ; \alpha}{}
        }
      }
    }
  }
 }
}
\]

Note that $\Pi_2$
introduces first the left-hand $\&$, then the right-hand $\&$. $\Pi_2$ 
is clearly equivalent by permutation of rules to the MALL proof
which introduces the two $\&$'s in the opposite order. There is an
analogous version of $\Pi'_2$ which introduces the two $\oplus$'s in
the opposite order. However, whereas $\Pi_2$ is essentially completely 
symmetric between the two $\&$'s, in the case of $\Pi'_2$ there is an
evident asymmetry between the left-hand $\&$, which is represented
implicitly by the $\oplus$ on the left-hand side of the sequent, and
the right-hand $\&$, which is represented in `double-negation' form as 
$\dnegwith$ on the right-hand side of the sequent. This asymmetry
makes a telling difference as regards the possibilities for forming
cuts, as we shall now see.

We now need an analogue  of $\Pi_3$. 
This should be a proof of a sequent
\[ \dnegwith \vdash A \oplus B ; \]
in which the last positive rule is the one introducing the $\oplus$ in 
the right-hand side of the sequent. This leads to the following proof 
$\Pi'_3$, in which we take $A = \shana$---this 
`double negation' is essential.
\[ \infer[\ShL]{\dnegwith \vdash \shana \oplus B ;}
 { \infer[\Foc]{\vdash \plusform , \shana \oplus B ;}
    {\infer[\PlusRl]{\vdash \plusform  ; \shana \oplus
        B}
     { \infer[\ShR]{\vdash \plusform ; \shana}
        {\infer[\ShL]{\shneg{\alpha} \vdash \plusform ;}
          {\infer[\Foc]{\vdash \alpha , \plusform ;}
            {\infer[\PlusRl]{\vdash \alpha ; \plusform}
             {\infer[\ShR]{\vdash \alpha ; \shneg{\alpha}}
              {\infer[\Foc]{\alpha \vdash \alpha ;}
               {\infer[\Ax]{\alpha \vdash ; \alpha}{}
               }
              }
             }
           }
         }
       }
     }
   }
 }
\]
Note the necessity to double-negate $\alpha$ in order to allow the
$\oplus$ which will end up on the left-hand side of the sequent to be introduced.
Note also that, crucially, because of the stoup constraint \emph{we cannot perform the
$\oplus$-introduction on the right hand side of the sequent as the
last rule in the proof}. This means that what we actually obtain combining $\Pi'_1$, 
$\Pi'_2$, $\Pi'_3$ is the analogous situation to case 7 in Table~1. If 
we worked in symmetric fashion from right to left, we would obtain a
situation analogous to case~1. However, \emph{no analogue of the problematic 
case 5 can be found in $\Mallf$}.

Although we have only considered a particular example, it should
illustrate the way in which $\Mallf$ tames MALL by putting sufficient
constraints on the control flow so that it admits a deterministic,
sequential scheduling.

It is precisely for this reason that $\Mallf$ is used in the work on
Ludics. While Ludics is a much richer and more subtle setting than
Blass games, it does have certain salient features in common. These
include the primacy of the additives, and also the fact that
interaction in Ludics is strictly polarized; in fact, the two players
strictly alternate, although either player can start. In Ludics,
control is retained over the scheduling by explicit use of the shift
operators (which in process terms can be seen as polarized versions of 
prefixing). So Ludics as a semantic universe corresponds to (and was
in some sense inspired by) the focussing version of MALL, and it is
for $\Mallf$ that a full completeness result is proved for Ludics in
\cite{Gir01}.

In fact, once one restricts attention to Linear Logic with Focussing,
or to the equivalent system $\mathsf{LLp}$ of \emph{polarized Linear
  Logic} \cite{LQT}, then more or less any form of sequential games can
be used to interpret the system without any problems of the kind
encountered with the full Classical system in section~3. Indeed,
Laurent has given a fully complete model for $\mathsf{LLp}$ using
Hyland-Ong style games \cite{Lau01}.
\section{The concurrent way out: Concurrent Games}
\label{cgsec}
When confronted with a mismatch between syntax and semantics, one may:
\begin{itemize}
\item modify the syntax, or
\item change the model.
\end{itemize}
Thus in the case of PCF, confronted with the non full abstraction of
the standard denotational model, one may extend PCF with parallel
constructs, or look for a more refined model which captures
sequentiality. In the case of Linear Logic, the movement is rather in
the opposite direction: starting from a sequential model, we can
either constrain the syntax to enforce sequentiality, or enrich the
model to allow concurrency to be represented directly. We now turn to
this latter possibility.

Neither game theory nor concurrency theory offers ready made solutions 
to this problem. While some suggestive remarks connecting games of
imperfect information with concurrent `teams' of players have been
made \cite{HS}, no precise, let alone tractable formalization has been
proposed so far as we are aware.

As we have seen, Blass games give primacy to the additive connectives
of Linear Logic. They give a direct and natural representation of the
idea that (occurrences of) $\&$ and $\oplus$ represent points in the
computation at which \emph{choices} have to be made. In the case of
$\&$, the choice is made by O, in the case of $\oplus$, by P. However, 
the account given of the multiplicative connectives in these games is
less satisfactory, as we have seen: the insistence on retaining a
global polarization, so that in each position it is exactly one
player's turn to move, leads to over-specification of the sequential
order of events, with bad formal consequences.

By contrast, we find an `authentic' semantic account of the
multiplicatives (and, to some extent, of the exponentials) in the
Geometry of Interaction \cite{Gi89,Gi90,Gi95,DR93,DR95}. Here the `local, asynchronous'
character of the multiplicatives, suggested by the intrinsic,
geometrical representation of multiplicative proofs as proof nets, is
turned into a concrete form of symbolic dynamics. The basic idea of
Geometry of Interaction is that
multiplicative proofs are represented by permutations acting on
structures of some kind.

\subsection{Background}\label{Linearlogicintro}

The original version of the Geometry of Interaction was developed by Girard
for the multiplicative fragment~\cite{Gir88}.  This is still the best
setting in which to explain the basic ideas on which the interpretation is
based.  

Consider then the multiplicative fragment, with
the restriction that the Axiom is only used for propositional atoms,
$\entails \alpha^{\perp}, \alpha$.  Now, if we look at the cut-free proofs,
for example of $\alpha \tensor \alpha \linimpl \alpha \tensor \alpha$, {\em
i.e. } of the sequent $\entails \alpha^{\perp} \parc \alpha^{\perp} ,\alpha
\tensor \alpha$, there are in fact just two, corresponding to the identity
and twist maps. 
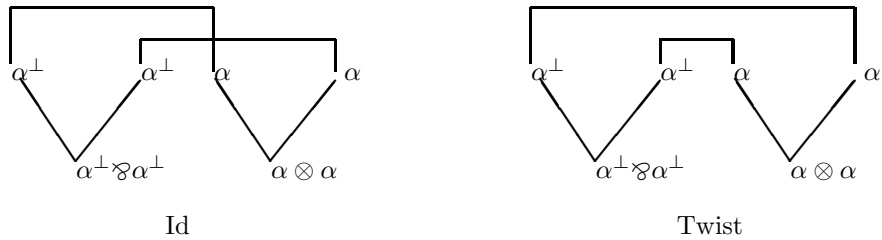
\setlength{\unitlength}{0.0085in}
\begin{center}
\begin{picture}(531,155)(100,650)
\thicklines
\put(260,705){\line( 4, 5){ 40}}
\put(260,705){\line(-2, 3){ 33.846}}
\put(140,705){\line( 4, 5){ 40}}
\put(140,705){\line(-2, 3){ 33.846}}
\put(100,765){\line( 0, 1){ 35}}
\put(100,800){\line( 1, 0){125}}
\put(225,800){\line( 0,-1){ 40}}
\put(180,765){\line( 0, 1){ 15}}
\put(180,780){\line( 1, 0){120}}
\put(300,780){\line( 0,-1){ 20}}
\put(460,705){\line(-2, 3){ 33.846}}
\put(460,705){\line( 4, 5){ 40}}
\put(580,705){\line(-2, 3){ 33.846}}
\put(580,705){\line( 4, 5){ 40}}
\put(500,765){\line( 0, 1){ 15}}
\put(500,780){\line( 1, 0){ 45}}
\put(545,780){\line( 0,-1){ 15}}
\put(420,765){\line( 0, 1){ 35}}
\put(420,800){\line( 1, 0){200}}
\put(620,800){\line( 0,-1){ 35}}
\put (505,660) {\makebox(0,0) [lb] {\raisebox{0pt}[0pt][0pt]{ Twist}}}
\put (190,660) {\makebox(0,0) [lb] {\raisebox{0pt}[0pt][0pt]{ Id}}}
\put (300,755) {\makebox(0,0) [lb] {\raisebox{0pt}[0pt][0pt]{ $\alpha$}}}
\put (220,755) {\makebox(0,0) [lb] {\raisebox{0pt}[0pt][0pt]{ $\alpha$}}}
\put (255,695) {\makebox(0,0) [lb] {\raisebox{0pt}[0pt][0pt]{ $\alpha \tensor \alpha$}}}
\put (175,755) {\makebox(0,0) [lb] {\raisebox{0pt}[0pt][0pt]{ $\alpha^{\perp}$}}}
\put (95,755) {\makebox(0,0) [lb] {\raisebox{0pt}[0pt][0pt]{ $\alpha^{\perp}$}}}
\put (135,695) {\makebox(0,0) [lb] {\raisebox{0pt}[0pt][0pt]{ $\alpha^{\perp} \parc \alpha^{\perp}$}}}
\put (455,695) {\makebox(0,0) [lb] {\raisebox{0pt}[0pt][0pt]{ $\alpha^{\perp} \parc \alpha^{\perp}$}}}
\put (415,755) {\makebox(0,0) [lb] {\raisebox{0pt}[0pt][0pt]{ $\alpha^{\perp}$}}}
\put (495,755) {\makebox(0,0) [lb] {\raisebox{0pt}[0pt][0pt]{ $\alpha^{\perp}$}}}
\put (575,695) {\makebox(0,0) [lb] {\raisebox{0pt}[0pt][0pt]{ $\alpha \tensor \alpha$}}}
\put (540,755) {\makebox(0,0) [lb] {\raisebox{0pt}[0pt][0pt]{ $\alpha$}}}
\put (620,755) {\makebox(0,0) [lb] {\raisebox{0pt}[0pt][0pt]{ $\alpha$}}}
\end{picture}
\label{example1}
\end{center}

As we see from these examples, cut-free proof nets in this fragment   have
the structure of a set of trees, one for each formula in the conclusion,
with the leaves connected up in pairs by the axiom links.  Moreover, the
structure of the trees is determined uniquely by the formulae in the
sequent (this is where the restriction on axioms is applied).
Hence, a complete invariant to distinguish the different cut-free proofs of
a given sequent is given by the information as to how the leaves are joined
up.
\\
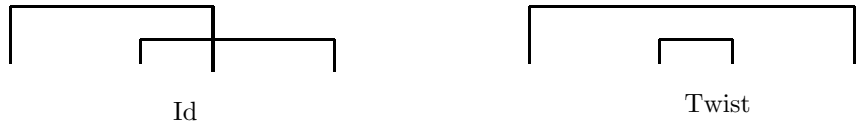
\begin{center}
\setlength{\unitlength}{0.0085in}
\begin{picture}(531,85)(95,720)
\thicklines
\put(100,765){\line( 0, 1){ 35}}
\put(100,800){\line( 1, 0){125}}
\put(225,800){\line( 0,-1){ 40}}
\put(180,765){\line( 0, 1){ 15}}
\put(180,780){\line( 1, 0){120}}
\put(300,780){\line( 0,-1){ 20}}
\put(500,765){\line( 0, 1){ 15}}
\put(500,780){\line( 1, 0){ 45}}
\put(545,780){\line( 0,-1){ 15}}
\put(420,765){\line( 0, 1){ 35}}
\put(420,800){\line( 1, 0){200}}
\put(620,800){\line( 0,-1){ 35}}
\put (510,735) {\makebox(0,0) [lb] {\raisebox{0pt}[0pt][0pt]{ Twist}}}
\put (195,730) {\makebox(0,0) [lb] {\raisebox{0pt}[0pt][0pt]{ Id}}}
\end{picture}
\end{center}
We can model this information by a permutation on the set of leaves
obtained as the product of the transpositions corresponding to the axiom
links.  Thus, 
\\
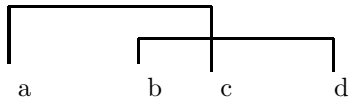
\begin{center}
\setlength{\unitlength}{0.0085in}
\begin{picture}(202,70)(300,735)
\thicklines
\put(370,765){\line( 0, 1){ 15}}
\put(370,780){\line( 1, 0){120}}
\put(490,780){\line( 0,-1){ 20}}
\put(290,765){\line( 0, 1){ 35}}
\put(290,800){\line( 1, 0){125}}
\put(415,800){\line( 0,-1){ 40}}
\put (485,745) {\makebox(0,0) [lb] {\raisebox{0pt}[0pt][0pt]{ d}}}
\put (415,745) {\makebox(0,0) [lb] {\raisebox{0pt}[0pt][0pt]{ c}}}
\put (370,745) {\makebox(0,0) [lb] {\raisebox{0pt}[0pt][0pt]{ b}}}
\put (290,745) {\makebox(0,0) [lb] {\raisebox{0pt}[0pt][0pt]{ a}}}
\end{picture}

\end{center}
corresponds to the permutation 
\[ \left( \begin{array}{cccc}
            a & b & c& d \\
            c & d & a& b
\end{array}
\right)
\]
Note that these transpositions are {\em disjoint}.  So, a cut-free proof is
represented by an {\em involution}, {\em i.e.} a self-inverse permutation.
This representation of cut-free proofs can be thought of as modelling the
`information flow' between the leaves in a dynamic fashion; think of
tokens travelling in both directions across the axiom links, as opposed to
modelling the linkage statically by a graph.  Note that we are using
functions to represent this information flow, but without input-output
bias, since the flow is bidirectional and symmetric.  

Returning to our example, consider performing cut-elimination on $twist
\circ twist$:
The proof net for ${\tt twist} \circ {\tt twist}$ before cut
elimination is:
\\
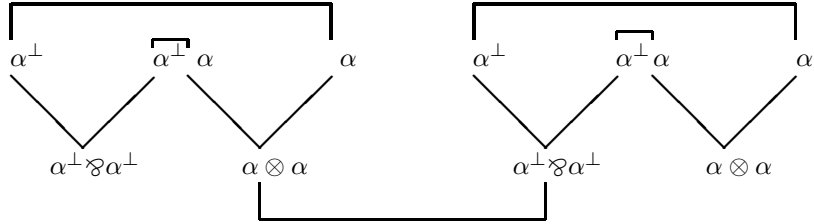
\begin{center}
\setlength{\unitlength}{0.0085in}
\begin{picture}(490,148)(99,653)
\thicklines
\put(390,752){\line( 1,-1){ 44.500}}
\put(434,707){\line( 1, 1){ 44.500}}
\put(500,752){\line( 1,-1){ 44.500}}
\put(544,707){\line( 1, 1){ 44.500}}
\put(214,752){\line( 1,-1){ 44.500}}
\put(258,707){\line( 1, 1){ 44.500}}
\put(105,752){\line( 1,-1){ 44.500}}
\put(149,707){\line( 1, 1){ 44}}
\put(105,774){\line( 0, 1){ 22}}
\put(105,796){\line( 1, 0){197}}
\put(302,796){\line( 0,-1){ 22}}
\put(192,769){\line( 0, 1){  5}}
\put(192,774){\line( 1, 0){ 22}}
\put(214,774){\line( 0,-1){  5}}
\put(390,774){\line( 0, 1){ 22}}
\put(390,796){\line( 1, 0){198}}
\put(588,796){\line( 0,-1){ 22}}
\put(478,774){\line( 0, 1){  5}}
\put(478,779){\line( 1, 0){ 22}}
\put(500,779){\line( 0,-1){  5}}
\put(258,686){\line( 0,-1){ 23}}
\put(258,663){\line( 1, 0){176}}
\put(434,663){\line( 0, 1){ 23}}
\put (143,691) {\makebox(0,0) [lb] {\raisebox{0pt}[0pt][0pt]{
$\hspace{-12pt}\alpha^{\perp} \parc \alpha^{\perp}$}}}
\put (253,691) {\makebox(0,0) [lb] {\raisebox{0pt}[0pt][0pt]{ $\hspace{-7pt}\alpha
\tensor \alpha$}}}
\put (428,691) {\makebox(0,0) [lb] {\raisebox{0pt}[0pt][0pt]{
$\hspace{-12pt}\alpha^{\perp} \parc \alpha^{\perp}$}}}
\put (539,691) {\makebox(0,0) [lb] {\raisebox{0pt}[0pt][0pt]{ $\hspace{-7pt}\alpha
\tensor \alpha$}}}
\put (99,758) {\makebox(0,0) [lb] {\raisebox{0pt}[0pt][0pt]{ $\alpha^{\perp}$}}}
\put (187,758) {\makebox(0,0) [lb] {\raisebox{0pt}[0pt][0pt]{ $\alpha^{\perp}$}}}
\put (214,758) {\makebox(0,0) [lb] {\raisebox{0pt}[0pt][0pt]{ $\alpha$}}}
\put (302,758) {\makebox(0,0) [lb] {\raisebox{0pt}[0pt][0pt]{ $\alpha$}}}
\put (384,758) {\makebox(0,0) [lb] {\raisebox{0pt}[0pt][0pt]{ $\alpha^{\perp}$}}}
\put (472,758) {\makebox(0,0) [lb] {\raisebox{0pt}[0pt][0pt]{ $\alpha^{\perp}$}}}
\put (495,758) {\makebox(0,0) [lb] {\raisebox{0pt}[0pt][0pt]{ $\alpha$}}}
\put (583,758) {\makebox(0,0) [lb] {\raisebox{0pt}[0pt][0pt]{ $\alpha$}}}
\end{picture}
\end{center}
The proof net for ${\tt twist} \circ {\tt twist}$ after one step of cut
elimination is:
\\
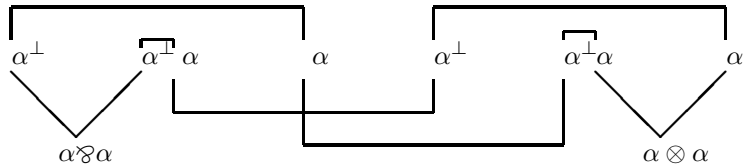
\begin{center}
\setlength{\unitlength}{0.0085in}
\begin{picture}(446,110)(95,700)
\thicklines
\put(280,760){\line( 0,-1){ 40}}
\put(280,720){\line( 1, 0){160}}
\put(440,720){\line( 0, 1){ 40}}
\put(200,760){\line( 0,-1){ 20}}
\put(200,740){\line( 1, 0){160}}
\put(360,740){\line( 0, 1){ 20}}
\put(440,785){\line( 0, 1){  5}}
\put(440,790){\line( 1, 0){ 20}}
\put(460,790){\line( 0,-1){  5}}
\put(360,785){\line( 0, 1){ 20}}
\put(360,805){\line( 1, 0){180}}
\put(540,805){\line( 0,-1){ 20}}
\put(180,780){\line( 0, 1){  5}}
\put(180,785){\line( 1, 0){ 20}}
\put(200,785){\line( 0,-1){  5}}
\put(100,785){\line( 0, 1){ 20}}
\put(100,805){\line( 1, 0){180}}
\put(280,805){\line( 0,-1){ 20}}
\put(100,765){\line( 1,-1){ 40}}
\put(140,725){\line( 1, 1){ 40}}
\put(460,765){\line( 1,-1){ 40}}
\put(500,725){\line( 1, 1){ 40}}
\put (535,770) {\makebox(0,0) [lb] {\raisebox{0pt}[0pt][0pt]{ $\alpha$}}}
\put (455,770) {\makebox(0,0) [lb] {\raisebox{0pt}[0pt][0pt]{ $\alpha$}}}
\put (435,770) {\makebox(0,0) [lb] {\raisebox{0pt}[0pt][0pt]{ $\alpha^{\perp}$}}}
\put (355,770) {\makebox(0,0) [lb] {\raisebox{0pt}[0pt][0pt]{ $\alpha^{\perp}$}}}
\put (280,770) {\makebox(0,0) [lb] {\raisebox{0pt}[0pt][0pt]{ $\alpha$}}}
\put (200,770) {\makebox(0,0) [lb] {\raisebox{0pt}[0pt][0pt]{ $\alpha$}}}
\put (175,770) {\makebox(0,0) [lb] {\raisebox{0pt}[0pt][0pt]{ $\alpha^{\perp}$}}}
\put (95,770) {\makebox(0,0) [lb] {\raisebox{0pt}[0pt][0pt]{ $\alpha^{\perp}$}}}
\put (495,710) {\makebox(0,0) [lb] {\raisebox{0pt}[0pt][0pt]{ $\hspace{-7pt}\alpha
\tensor \alpha $}}}
\put (135,710) {\makebox(0,0) [lb] {\raisebox{0pt}[0pt][0pt]{ $\hspace{-7pt}\alpha
\parc \alpha$}}}
\end{picture}
\end{center}
Generally, in this fragment, we can apply this `decomposition rule'
repeatedly for tensors cut against par until all cuts are between axiom
links.  We can say that the whole purpose of these transformations is
to match up the corresponding axiom links correctly; the `real'
information flow is then accomplished by the axiom reductions:
\\
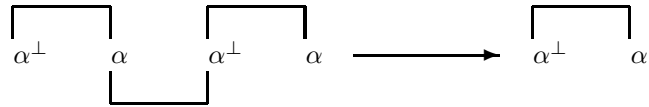
\begin{center}
\setlength{\unitlength}{0.0085in}
\begin{picture}(386,75)(135,695)
\thicklines
\put(140,745){\line( 0, 1){ 20}}
\put(140,765){\line( 1, 0){ 60}}
\put(200,765){\line( 0,-1){ 20}}
\put(200,725){\line( 0,-1){ 20}}
\put(200,705){\line( 1, 0){ 60}}
\put(260,705){\line( 0, 1){ 20}}
\put(260,745){\line( 0, 1){ 20}}
\put(260,765){\line( 1, 0){ 60}}
\put(320,765){\line( 0,-1){ 20}}
\put(460,745){\line( 0, 1){ 20}}
\put(460,765){\line( 1, 0){ 60}}
\put(520,765){\line( 0,-1){ 20}}
\put(350,735){\vector( 1, 0){ 90}}
\put (135,730) {\makebox(0,0) [lb] {\raisebox{0pt}[0pt][0pt]{ $\alpha^{\perp}$}}}
\put (195,730) {\makebox(0,0) [lb] {\raisebox{0pt}[0pt][0pt]{ $\alpha$}}}
\put (255,730) {\makebox(0,0) [lb] {\raisebox{0pt}[0pt][0pt]{ $\alpha^{\perp}$}}}
\put (315,730) {\makebox(0,0) [lb] {\raisebox{0pt}[0pt][0pt]{ $\alpha$}}}
\put (455,730) {\makebox(0,0) [lb] {\raisebox{0pt}[0pt][0pt]{ $\alpha^{\perp}$}}}
\put (515,730) {\makebox(0,0) [lb] {\raisebox{0pt}[0pt][0pt]{ $\alpha$}}}
\end{picture}
\end{center}
or more generally,
\\
\begin{center}
\setlength{\unitlength}{0.0085in}
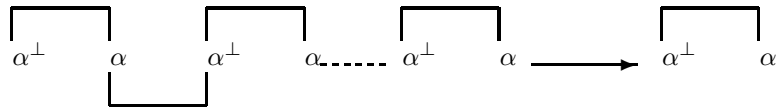
\begin{picture}(466,75)(95,710)
\thicklines
\put(340,760){\line( 0, 1){ 20}}
\put(340,780){\line( 1, 0){ 60}}
\put(400,780){\line( 0,-1){ 20}}
\put(100,760){\line( 0, 1){ 20}}
\put(100,780){\line( 1, 0){ 60}}
\put(160,780){\line( 0,-1){ 20}}
\put(160,740){\line( 0,-1){ 20}}
\put(160,720){\line( 1, 0){ 60}}
\put(220,720){\line( 0, 1){ 20}}
\put(220,760){\line( 0, 1){ 20}}
\put(220,780){\line( 1, 0){ 60}}
\put(280,780){\line( 0,-1){ 20}}
\put(500,760){\line( 0, 1){ 20}}
\put(500,780){\line( 1, 0){ 60}}
\put(560,780){\line( 0,-1){ 20}}
\multiput(290,745)(7.27273,0.00000){6}{\line( 1, 0){  3.636}}
\put(420,745){\vector( 1, 0){ 65}}
\put (95,745) {\makebox(0,0) [lb] {\raisebox{0pt}[0pt][0pt]{ $\alpha^{\perp}$}}}
\put (155,745) {\makebox(0,0) [lb] {\raisebox{0pt}[0pt][0pt]{ $\alpha$}}}
\put (215,745) {\makebox(0,0) [lb] {\raisebox{0pt}[0pt][0pt]{ $\alpha^{\perp}$}}}
\put (275,745) {\makebox(0,0) [lb] {\raisebox{0pt}[0pt][0pt]{ $\alpha$}}}
\put (335,745) {\makebox(0,0) [lb] {\raisebox{0pt}[0pt][0pt]{ $\alpha^{\perp}$}}}
\put (395,745) {\makebox(0,0) [lb] {\raisebox{0pt}[0pt][0pt]{ $\alpha$}}}
\put (495,745) {\makebox(0,0) [lb] {\raisebox{0pt}[0pt][0pt]{ $\alpha^{\perp}$}}}
\put (555,745) {\makebox(0,0) [lb] {\raisebox{0pt}[0pt][0pt]{ $\alpha$}}}
\end{picture}
\end{center}
The idea, as with cut-free proofs, is to model these transformations
dynamically, by the flow of information tokens, rather than by graph
rewriting.  

An interpretation of the multiplicative fragment can be given using just
permutations on finite sets, as described in \cite{Gir88}.  
The task of \emph{characterizing} the space of proofs is then to pick out
exactly those permutations which can arise as the denotation of (cut-free)
proof nets, and moreover to do so in a compositional, syntax-free
fashion. This is the goal of \emph{full completeness theorems}, to be
discussed in the next section. The main point to note for now is that
this can indeed be done using suitable uniformity and preservation
properties in a number of semantic settings; this is the content of the
various full completeness theorems for Multiplicative Linear Logic
which have appeared over the past few years, starting with \cite{gfc}
(see e.g. \cite{HO93,BS,Loa94,Hag00a}).

The question is, how can this view of the multiplicatives, which has
no scope for expressing causal dependencies (this is exactly the sense in 
which it is asynchronous) be reconciled with the additives, which as
we have seen are essentially concerned with choice and
causality?\footnote{The extension of Geometry of Interaction to the
  additives in \cite{Gi95}, while containing much of interest, is
  generally agreed not to offer a fully satisfactory treatment.}

We can in fact characterize the behavioural features of the various
levels of connectives in Linear Logic using the concepts of Net Theory 
\cite{Pet77}:
\begin{itemize}
\item the multiplicatives express \emph{concurrency} (i.e. pure causal
  independence)
\item the additives express \emph{conflict} and \emph{causality}
  (i.e. choice and sequencing).
\end{itemize}
(This is of course just the `true concurrency' perspective on our
previous alignment of the additives with the dynamic operations of
process algebra, and the multiplicatives with the static operations.)
What we are looking for is a setting in which all these features can
be represented.

Of course, Net Theory itself offers such a setting. The problem with
using Net Theory as a formal basis for our semantic representation is
that, as we have seen, we need \emph{two levels of description}: the
formulas \emph{and} the proofs (or: the games \emph{and} the
strategies). While it was a relatively simple matter to describe
strategies as subtrees of synchronization trees, it does not seem to
be so easy to find a good, tractable description of strategies over
games represented as Petri Nets. For this reason, we shall pursue a
different approach.

In keeping with the discussion so far, we are looking for a semantic
representation such that both:
\begin{itemize}
\item the account of the additives given in Blass games, and
\item the account of the multiplicatives given in Geometry of
  Interaction
\end{itemize}
can be faithfully represented.
By being a little more abstract, an elegant account meeting all these
desiderata can be found.

\subsection{A domain-theoretic formalization of games and strategies}
Let us begin by revisiting Blass games,
with the idea that game 
trees can be viewed as partial orders, in which $x \leq y$ 
means that the position $y$ can be reached from the position $x$ by 
playing some additional moves. This is a natural ``information 
ordering'' as in Domain theory \cite{handbookchap}; it is the prefix
ordering on the plays or paths through the game tree. If we add ``limit points'' 
corresponding to the infinite branches in the game tree, we obtain a  
complete partial order $D$.

Viewed in these terms, the construction of sums and products of games 
as in Blass games can be described as \emph{lifted sums} as far as the 
underlying domains of positions are concerned:
\[ D(\coprod_{i \in I} A_{i}) = D(\prod_{i \in I} A_{i}) = (\biguplus_{i \in I} 
D(A_{i}))_{\bot} . \]
Here $\biguplus_{i \in I} D_i$ is the disjoint union of the partial
orders $D_i$; while $D_\bot$ is the lift of $D$, i.e. the result of
adding a new bottom element to $D$. It is important to understand that 
this partial order representation is entirely equivalent to that given 
using deterministic synchronization trees in Section~2. The root of
the game tree for $G$ is represented by the bottom element of $D(G)$,
corresponding to the position in which no moves have yet been
played. The elements of $D(G)$ immediately above $\bot_D$ will be of
the form $\mathsf{in}_i(\bot_{i})$, where $\bot_i$ is the bottom
element of $G_i$. Such an element corresponds to the position where
the opening move $i$ has been made, and we are now at the root of the
sub-tree $G_i$. As we go up in the ordering, this will correspond to
the further development of the play. 

Importantly, the partial order representation of game trees admits an 
elegant representation of \emph{strategies},
as functions on these domains of 
positions: $f : D \rightarrow D$, where $f(x)$ is the position 
obtained from $x$ by extending it with whatever moves the strategy 
makes in that position. It is then immediate that $f(x) \sqsupseteq 
x$. Moreover, those positions where $f$ has no moves to make (e.g. 
because `it is not its turn') are exactly the \emph{fixpoints} of 
$f$. In the usual way, we require computationally reasonable 
strategies to be monotonic and continuous. Finally, as a useful 
normalizing condition, we require strategies to be 
\emph{idempotent}: $f^{2} = f$. To understand this, consider $f$ 
applied to $f(x)$. The only moves made in $f(x)$ which were not 
already made in $x$ are those made by $f$ itself: $f(x)$ contains no 
more information supplied by the Opponent (i.e. the environment) than 
$x$ did. Hence anything $f$ decides to do at $f(x)$ it should have 
already been able to decide to do at $x$, and we require that 
$f(f(x)) = f(x)$. Of course, this allows several moves to be made in a 
block by a player. This possibility already exists in Blass games, 
e.g. Opponent must move twice initially in any game of the form
\[ (A \sqcap B) \sqcap (C \sqcap D) , \]
(first to resolve the choice between $(A \sqcap B)$ and $(C \sqcap
D)$, then to resolve the initial choice offered in the chosen sub-game).
An important point is that strategies may not be well-defined at all 
positions. In general there are some positions that can never be 
reached by following that strategy. To mesh with the requirement that 
strategies are increasing functions, we adjoin a top element to the 
domain of positions $D$, writing this as $D^{\top}$. We represent 
$f$ being undefined at $x$ by $f(x) = \top$.
Note that in this context $\top$ is really being used to represent
\emph{undefinedness} or partiality of the closure operator,
i.e. \emph{divergence}. On the other hand, being stuck at a
(non-maximal) fixpoint can be userstood as \emph{deadlock}. Thus a
natural distinction appears in this setting between deadlock and divergence.

We recall that a \emph{closure operator} on a partial order is an
idempotent, increasing, monotone endofunction.
So our discussion can be summarized by saying that strategies (for either player) are represented as 
continuous closure operators on $D^{\top}$, which under modest 
assumptions on $D$ (bounded completeness) is a complete lattice.
\paragraph{Notation} We write $\sigma : D^{\top}$ to signify that
$\sigma$ is a continuous closure on $D^{\top}$.

\noindent We can completely specify a game as a structure $(D, S^{P}, S^{O})$, 
where $D$ is the domain of positions, $S^P$ is the set of legal 
strategies for P, and $S^{O}$ is the set of legal 
counter-strategies, i.e. strategies for O. 

\paragraph{Remark} We remark in passing that closure operators have been used to give a
denotational semantics to (deterministic) concurrent constraint
programming \cite{SRP91}. The basic dynamics of this model is very much akin 
to what we are doing here, just as the basic dynamics of process
calculi is akin to the structure of sequential games. The added
ingredient in each case is the extra structure implied by
distinguishing explicitly between System and Environment, and imposing 
a typed framework by distinguishing between the game and the strategy.
In the work on concurrent constraint programming, closure operators
are studied in terms of their sets of fixpoints. This is very
convenient from a denotational point of view, but if the dynamics are
being studied, then the properties of the closures \textit{qua} functions are
significant, and we prefer to study them in these terms.

We can define the strategies for either player inductively on
(well-founded) Blass games, just as we did using the synchronization
tree representation.
For products, we define
$$S^{P}(\prod_{i \in I} A_{i}) = \{ \tuple{\sigma_{i}}{i \in I}
\mid \forall i\in I. \, \sigma_{i} \in S^{P}(A_{i})\} $$
where
$$ \begin{array}{lcl}
\tuple{\sigma_{i}}{i \in I} (\bot ) & = & \bot \\
\tuple{\sigma_{i}}{i \in I} (\inj_{i}(x) ) & = & 
\inj_{i}(\sigma_{i} (x)) .
\end{array} $$
This corresponds exactly to the idea that Player must first wait for 
Opponent to choose an $i \in I$, and then plays according to some 
strategy for $A_{i}$.
$$ S^{P}(\coprod_{i \in I} A_{i}) = \{ \inj_{i}(\sigma) \mid 
i\in I \wedge \sigma \in S^{P}(A_{i})\} $$
where
$$ \begin{array}{lcl}
\inj_{i}(\sigma )(\bot ) & = & \inj_{i}(\sigma (\bot )) \\
\inj_{i}(\sigma )(\inj_{i}(x) ) & = & \inj_{i}(\sigma (x )) \\
\inj_{i}(\sigma )(\inj_{j}(x) ) & = &  \top , \;\; (i \not= j).
\end{array} $$
Again, this corresponds to the idea that a strategy for Opponent will 
firstly choose some $i$, then play according to a strategy for 
Opponent in $A_{i}$. Note that the last case above covers 
`unreachable states' for $\inj_{i}(\sigma)$.

We can also define the duality directly in terms of its effect on the
set of strategies for either player: the reversal of r\^{o}les means
that strategies for P turn into strategies for O, and vice versa:
\[ (D, S^P , S^O )^{\perp} = (D, S^O , S^P ) . \]
Clearly with this definition $G^{\perp \perp} = G$.
Also, we define $S^{O}(G) = S^{P}(G^{\perp})$.

For the case of well-founded Blass games, we recover exactly the same
class of strategies as in our earlier definition in terms of
synchronization trees.
For such games, $D$ is the domain of finite 
and infinite sequences under the prefix ordering corresponding to the 
paths through the game tree, and the conventions about who is to play 
are captured by the fact that, for all $x \in D$, either $\sigma (x) = 
x$ for all $\sigma \in S^{P}$ (it is Opponent's turn to move), or $\tau (x) = 
x$ for all $\tau \in S^{O}$ (it is Player's turn). However, this 
is a very special case of our general setting; and we will overcome 
the problems with Blass games precisely by allowing situations in 
which \emph{both} players can move.

To do this, we shall interpret the game boards for the 
multiplicatives differently to Blass: by a true concurrency rather 
than an interleaving representation. In our setting,  this is
simply a matter of defining
$$ D(A \otimes B) = D(A \llpar B) = D(A) \times D(B), $$
the cartesian product of domains.
It is important to understand how this relates to our earlier
discussion of true concurrency.
A typical situation in which there are two causally independent events
$a$ and $b$ might be represented by a game $D \times E$, where say $a$ 
corresponds to the increase in information $\bot \sqsubseteq d$ in $D$, while
$b$ is represented by the increase in information $\bot \sqsubseteq e$ 
in $E$. Then the performance of these two actions \emph{in either
  order}
\[ (\bot , \bot ) \sqsubseteq (d, \bot ) \sqsubseteq (d, e) \]
corresponding to the trace $ab$, or
\[ (\bot , \bot ) \sqsubseteq (\bot , e) \sqsubseteq (d, e) \]
corresponding to $ba$, will indeed be identified, with the single
outcome $(d, e)$. Indeed, cartesian product of event structures is
standardly used to model parallel composition in `partial order' or
`true concurrency' semantics \cite{WN95}.

For the strategies for tensor, we define
\[ S^{P}(A \otimes B) = \{ \sigma \times \tau \mid \sigma \in S^{P}(A) 
\wedge
\tau \in S^{P}(B) \} \]
where $\sigma \times \tau (x, y) = (\sigma (x), \tau (y))$. (This is 
really smash product with respect to $\top$; if either $\sigma (x) = 
\top$ or $\tau (y) = \top$, then the result is $\top$.)
This exactly captures the idea of informational independence between 
Player's actions in $A$ and in $B$ (cf. \cite{gfc}). How Player moves 
in $A$ depends only on the information available in $A$, and 
similarly for $B$.

In order to define the counter-strategies for the Tensor (and hence 
the strategies for Par and Linear implication, and eventually the 
morphisms in the category of concurrent games), we introduce the most 
important feature of our formalization: the elegant treatment it 
affords of composition of strategies. Suppose firstly that $\sigma \in 
S^{P}$ and $\tau \in S^{O}$ in a game $(D, S^{P}, S^{O})$. How do we 
play $\sigma$ off against $\tau$? We define
$$\begin{array}{rcl}
\eval{\sigma}{\tau}&=& \mathbf{Y} (\sigma \circ 
\tau ) = \bigvee_{k \in \omega} (\sigma \circ \tau)^{k} (\bot)\\
&=& \bigvee_{k \in \omega} (\tau \circ \sigma)^{k} (\bot )
= \mathbf{Y} (\tau \circ \sigma ) .
\end{array}$$
The fact that these two least fixpoints coincide follows easily from 
the fact that $\sigma$ and $\tau$ are continuous and increasing; in particular, 
this is a special case of the construction of the \emph{join} of two 
closure operators. Thus $\eval{\sigma}{\tau} \in D$ is 
the position we reach as a result of playing $\sigma$ against 
$\tau$. The equality of the two formulas above also shows that this is 
independent of all questions about `who starts'.

Now given closure operators $\sigma$ on $(D \times E)^{\top}$ and 
$\tau$ on $(E \times F)^{\top}$, we want to `compose' them to get a 
closure $\sigma ; \tau$ on $(D \times F)^{\top}$. We define this as 
follows:
$$ \sigma ; \tau (x, z) = (\pi_{1} \circ \sigma (x, y), \pi_{2} \circ 
\tau (y, z)) $$
where
$$y = \eval{\pi_{2} \circ \sigma (x, {-})}{\pi_{1} \circ \tau ({-}, z)}.$$
That is, given input in $D$ and $F$, we play $\sigma$ and $\tau$ off 
against each other in $E$ relative to this input, 
and obtain their external response taking 
into account their interaction with each other.

In particular, if $\sigma$ is a closure on $\up{(D \times E)}$, it 
induces an `action' taking closures on $\up{D}$ to closures on 
$\up{E}$, $\alpha \mapsto \alpha ; \sigma$, and a `coaction' 
taking closures on $\up{E}$ back to closures on $\up{D}$, 
$\beta \mapsto \sigma ; \beta$.
E.g. $\sigma ; \beta (x) = \pi_{1} \circ \sigma (x, y)$, where $y = 
\eval{\pi_{2} \circ \sigma (x, {-})}{\beta}$.

Now we can define
$$\begin{array}{rl}
S^{O}(A \otimes B) = \{ \sigma : (D(A) \times D(B))^{\top} 
\mid & \forall \alpha \in S^{P}(A). \, \alpha ; \sigma \in S^{O}(B) \\
\wedge & \forall \beta \in S^{P}(B). \, \sigma ; \beta \in S^{O}(A) 
\} .\end{array}$$
Again, by De Morgan duality,
$$ A \llpar B = (A^{\bot} \otimes B^{\bot})^{\bot}, \;\;\;
A \linimpl B = A^{\bot} \llpar B . $$
In particular,
$$\begin{array}{rl}
S^{P}(A \linimpl B) = \{ \sigma : (D(A) \times D(B))^{\top} 
\mid & \forall \alpha \in S^{P}(A). \, \alpha ; \sigma \in S^{P}(B) \\
\wedge & \forall \beta \in S^{O}(B). \, \sigma ; \beta \in S^{O}(A)
\} .\end{array}$$
This is a symmetric, `classical' version of the familiar logical 
relations condition: $\sigma$'s action carries P strategies on $A$ to 
P strategies on $B$, and its coaction carries counter-strategies on 
$B$ to counter-strategies on $A$. 

\subsection{Connection with Geometry of Interaction}
As to the issue of faithfully reflecting geometry of interaction ideas 
in modelling the multiplicatives, we recall that the basic idea in
geometry of interaction is that
multiplicative proofs are modelled by permutations.
Note that in any symmetric monoidal category, the symmetric group
$S(n)$ acts in a canonical way on the tensor power
\[ \bigotimes^n A \; = \; \underbrace{A \otimes \ldots \otimes A}_n
. \]
In the original form of Geometry of Interaction \cite{Gi89} (dubbed `particle
style GoI' in \cite{Abr96}), the monoidal structure used for the
representation is coproduct (disjoint union). In our setting, we can
use products to exactly the same effect. Thus in our example of the
sequent
\[ \entails \alpha^{\perp} \parc \alpha^{\perp} ,\alpha
\tensor \alpha \]
the two proofs can be represented as maps on a product
\[ D \times D \times D \times D \]
by
\[ (a, b, c, d) \mapsto (c, d, a, b) \]
for the identity, and
\[ (a, b, c, d) \mapsto (d, c, b, a) \]
for the twist.

This idea, although basically correct, requires a slight elaboration
in order to mesh with our approach of using closure operators to model 
strategies. We turn permutations $\sigma$ as above into closure
operators $\gamma$ by defining
\[ \gamma (x) = x \vee \sigma (x) . \]
Thus for example the permutation corresponding to the identity as
above will give rise to the closure operator
\[ (a, b, c, d) \mapsto (a \vee c, b \vee d, c \vee a, d \vee b). \]
The axiom $\entails \alpha^{\bot}, \alpha$ is interpreted by the twist 
map on $D \times D$
\[ (x, y) \mapsto (y, x) \]
and then by the closure
\[ (x, y) \mapsto (x \vee y, x \vee y) \]
which forces the pair $(x, y)$  to the least element above it in the 
product $D \times D$ which lies on the diagonal (and to $\top$ if $x$
and $y$ have no
upper bound in $D$).

In this fashion, we can claim to have achieved our objective of
faithfully embedding both the Blass games treatment of the additives,
and the GoI treatment of the multiplicatives, within our model.

\subsection{Connection with `New Foundations'}
It should be noted that mathematically, the concurrent games model is
merely a rephrasing of the `New Foundations for the Geometry of
Interaction' introduced by the present author and Radha Jagadeesan in
\cite{nfgoi}. As demonstrated in \cite{Abr96,ahs}, this and the original form of
geometry of interaction in \cite{Gi89} are indeed both instances of a
single general scheme, as far as the interpretation of the
multiplicatives and exponentials are concerned. However, as already
emphasized in \cite{nfgoi}, the New Foundations setting has the
representational capacity to capture the causality inherent in the
additives, which the `particle style' GoI of \cite{Gi89} does not. The
advantage of the presentation of concurrent games in terms of closure
operators is that the connection with traditional forms of game
semantics, and Blass games in particular, becomes much
clearer. However, the New Foundations presentation remains important,
particularly since we can use it to pick out those strategies
satisfying significant domain-theoretic properties such as stability
and sequentiality, which turn out to be important for full
completeness, i.e. for characterizing which strategies arise from
proofs. These properties apply to the underlying NFGoI functions $f$, rather than 
to the closure operators $\sigma$ they induce via
\[ \sigma (x) = x \vee f(x). \]
For example, as already explained in the previous section, we
interpret axioms by the closure
\[ (x, y) \mapsto (x \vee y, x \vee y) \]
which is certainly not a stable function; however, its underlying
NFGoI function is the twist map, which  of course \emph{is} stable.

It is also worth pointing out to readers with a concurrency
background that the New Foundations interpetation can be understood
operationally in terms of a simple dataflow model, as described in
\cite{nfgoi}. We have already pointed out the connection to
concurrent constraints, which can be seen as a generalization of
dataflow. (In fact, the first use of closure operators in the
semantics of concurrency was for the
semantics of a dataflow language with logic variables \cite{JPP89}).

\subsection{Composition revisited}
We now illustrate how the concurrent games model overcomes the problem 
with composition which arises with Blass games, with reference to the
example discussed in section~\ref{syntpers}.

For illustration, we fix the interpretation of $\alpha$ to be the game 
$\Bool = I \oplus I$, where
\[ I = (\mathbf{1}, \{ \mathsf{id}_{\mathsf{1}} \} ,  \{
  \mathsf{id}_{\mathsf{1}} \} ) \]
the one-point domain in which no moves are possible (i.e. the sole
point is the root of the game tree), and in which the identity
function is a valid strategy for both P and O.
Note that $\Bool$ can be seen as a type of `booleans':
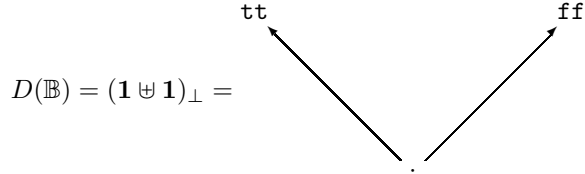
\[ D(\Bool ) = (\mathbf{1} \uplus \mathbf{1})_{\bot}
= \begin{diagram}
\mathtt{tt} & & & & \mathtt{ff} \\
& \luTo & & \ruTo & \\
& & \cdot & & \\
\end{diagram} \]
the flat domain of booleans, with
\[ S^{P}(\Bool ) = \{ \mathsf{in}_{\mathtt{tt}}, \mathsf{in}_{\mathtt{ff}} \} 
\qquad S^{O}(\Bool ) = \{ \mathsf{id}_{\Bool} \}  \]
where for $b \in \{ \mathtt{tt}, \mathtt{ff} \}$
\[ \inj_b (\bot ) = \inj_b (b) = b, \quad \inj_b (b') = \top \quad (b \neq
b' ) .  \]
We can now compute the underlying domains of the sequents:
\[ \begin{array}{lcl}
D(\alpha^{\bot} \oplus \alpha^{\bot}, \alpha \oplus \alpha ) & = & 
D((\alpha^{\bot} \oplus \alpha^{\bot}) \llpar (\alpha \oplus \alpha )) \\
& = & (\Bool \uplus \Bool )_{\bot} \times (\Bool \uplus \Bool )_{\bot} 
\\
& = & D(\alpha^{\bot} \& \alpha^{\bot}, \alpha \& \alpha ).
\end{array} \]
Although these domains are identical, if we take account of the
polarities on the additive connectives, which are reflected in the
definitions of the sets of strategies for the corresponding games, we can distinguish between them. For
example,
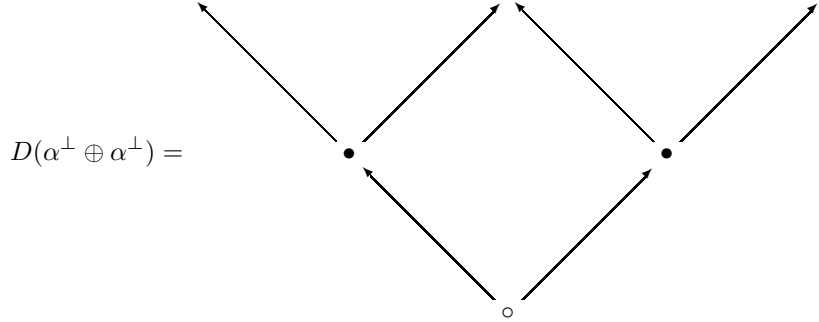
\[ D(\alpha^{\bot} \oplus \alpha^{\bot}) = 
\begin{diagram}
 & & &   &  &   & & &  \\
& \luTo & & \ruTo &    & \luTo & & \ruTo & \\
& & \bullet & &    & & \bullet & & \\
& & & \luTo &    & \ruTo & & & \\
& & & &   \circ   & & & & \\
\end{diagram} \]
Here the nodes marked $\bullet$ are where O is to move, and hence
must be fixpoints for all
P-strategies; and conversely for the $\circ$-node.

\noindent We now consider the interpretations of the three proofs discussed in
section~\ref{syntpers}. Firstly, $c = \llbracket \Pi_1 \rrbracket$ has
the definition
\[ c(\bot , \bot ) = (\intt (\bot ), \intt (\bot )) \]
\[ c(\intt (x), \intt (y)) = (\intt (x \vee y), \intt (x \vee y)) \]
\[ c(\inff (x), y) = c(x, \inff (y)) = \top . \]
Note that $\Pi_3$, the proof which introduces the two occurrences of
$\oplus$ in the opposite order, has \emph{the same denotation}
$\llbracket \Pi_3 \rrbracket = e = c$; in both cases, the two
introductions (or: choices at the $\oplus$-nodes) are performed concurrently.

Finally $d = \llbracket \Pi_2 \rrbracket$  has the definition
\[ d(x, \bot ) = (x, \bot ) \qquad d(\bot , y) = (\bot , y) \]
\[ d(\inj_{i}(x), \inj_{j}(y)) = (\inj_{i}(x \vee y), \inj_{j}(x \vee
y)) . \]
Again, this has the same denotation as the proof which introduces the
two $\&$'s in the opposite order. (However, it should be noted that in 
general, strategies must be taken modulo partial equivalence relations
as in \cite{AM99} in order to obtain the unicity properties for product
and coproduct (i.e. the commutative conversions in sequent calculus
terms, or $\eta$-conversions in $\lambda$-calculus terminology). Interestingly, partial equivalence relations are also
needed in Ludics \cite{Gir01}).

Now one can check, unfolding the definitions, that
\[ c; (d;e) = c = e = (c;d) ; e . \]
For example, the fixpoint in $c;d$ converges after two iterations,
with the resulting behaviour
\[ (c;d)(\bot , \bot ) = (\intt (\bot ), \bot ) \]
\[ (c;d) (\intt (x), \inj_{j} (y)) = (\intt (x \vee y), \inj_{j}(x \vee
y)) . \]
This strategy, of type 
\[ \girard \alpha^{\bot} \oplus \alpha^{\bot}, \alpha \&
\alpha \]
exemplifies how \emph{global polarization no longer holds} in the concurrent
games model. \emph{Both} P \emph{and} O can move initially, P to make the choice at
the $\oplus$, O to make the choice at the $\&$. After these choices
have been made (determining an axiom link in the corresponding proof
net) a copy-cat is then played between the chosen occurrences of the
atom. See \cite{nfgoi} for many examples of such fixpoint computations, and 
  a detailed analysis of the correspondence with cut elimination.

Thus when we compose in either order, $(c;d) ;e$ or $c;(d;e)$, the
result is the same; P is no longer blocked by polarization constraints 
from making the choices at \emph{both} occurrences of $\oplus$---truly
concurrently.

\subsection{A more concrete presentation}
The foregoing mathematical development may  have seemed 
abstract, and hard to relate to one's intuitions.
Let us start again, and give an essentially equivalent, but more concrete
presentation.

Our notion of concurrent games can be thought of as played between
two `teams' of players, distributed around the `game board', in an
aynchronous fashion.

The basic action in a game is \emph{making a move}, which is chosen
from among several possible alternatives. The information content of the
move is the \emph{choice} or decision as to which move to play.

We fix a set $C$ of `cells' as labels for the various choices or decisions
that have to be made during the course of a play of the game, and a set
$V$ of `values', enumerating the possible choices. So the `event'
$(c, v)$ means that alternative $v$ was chosen at cell $c$.
A state of the game board---a position of the game---is then represented
by a partial function from cells to values
\[ \{ (c_1 , v_1 ), \ldots , (c_k , v_k ) \} \]
(so there is \emph{no} information about the order in which decisions were
taken).

We can take $C = V = \Nat$  for a `universal' choice of game board,
into which any other (subject to countability) can be embedded. 
The positions are then partial
functions on natural numbers, which under inclusion form one of the
basic examples of a \emph{domain} \cite{handbookchap}. Moreover, as play 
progresses, and more moves are made,
the position \emph{increases} in the natural information ordering:
\[ \{ (c_1 , v_1 ) , (c_2 , v_2 ) \} \stackrel{(c_3 , v_3 )}{\longrightarrow}
\{ (c_1 , v_1 ) , (c_2 , v_2 ), (c_3 , v_3 ) \} . \]
Call this universal game board $\GG$.

A \emph{strategy} is a function
\[ f : \GG \longrightarrow \GG . \]
For the reasons already explained in the previous section, this function 
should be monotonic and continuous, increasing ($x \subseteq f(x)$), and
idempotent ($f(f(x)) = f(x)$).

What about unreachable positions? We need an `error element' $\top$
for situations when the strategy is applied to a position which could
never have been reached if we had been following that strategy. To
preserve the increasingness property, this error element must be
top in the information ordering. So a strategy is a continuous closure
operator on the complete lattice $\GG^{\top}$ obtained by adjoining a top
element to $\GG$.
The set of all such closure operators
\[ \Strat = \mathsf{Cl}({\GG}^{\top} ) \]
is our universal space of strategies.

Various sub-classes of $\Strat$ are important. In particular, there
are constraints on strategies which correspond to important computational
properties. One of these is sequentiality \cite{Cur86}.

A closure operator $\sigma \in \Strat$ has an \emph{output function}
$f : \GG \rightarrow \GG$ if
\[ \forall x \in \mathtt{dom} (\sigma ). \, \sigma (x) = f(x) \vee x \]
where
\[ \mathtt{dom} (\sigma ) = \{ x \in \GG \mid \sigma (x) \neq \top \} . \]
We recall the domain-theoretic conditions of \emph{stability} and
\emph{sequentiality} \cite{amadio-curien}. These make sense because our underlying domain
$\GG$ is simple and concrete---it is in fact a `concrete domain' in the 
technical sense of Kahn and Plotkin \cite{KP78}. Indeed, because it is simply
a (countable) product of flat domains (in fact, $\GG = {\Nat_{\bot}}^{\omega}$),
we can use Vuillemin's original definition of sequentiality \cite{Vui73}: 
a function
$f : \GG \rightarrow \GG$ is sequential if:
\[ \forall c \in C. \, \forall x \in \GG . \, f(x)(c){\uparrow} \;\; \Rightarrow
\;\; \forall y \supseteq x. \, f(y)(c) {\uparrow} \]
\[ \vee \;\; (\exists c'. \, x(c'){\uparrow} \wedge \forall y \supseteq x. \,
f(y)(c){\downarrow} \; \Rightarrow \; y(c'){\downarrow} ) . \]
This says that there is some particular cell in the input which \emph{must}
get  filled before we can fill the given output cell.
Classically, this condition excludes the parallel-or function, Berry's
`Gustave function', etc. \cite{Vui73,KP78,Cur86}.

We take the `constraint' of having
a sequential output function as picking out those strategies which are
to be regarded as computing in a deterministic fashion, and hence
sequentially implementable.
In support of this contention, we have the fact that, when we make our universe
of sequential strategies into a model for a sequential functional language
such as PCF (an elegant way of doing this is to use realizability ideas 
\cite{Abr99}), then it is fully abstract, and even universal if we restrict 
to effective strategies.

It is interesting to contrast this with Longley's use of realizability 
to characterize the \emph{strongly stable model} of Bucciarelli and
Ehrhard \cite{Lon}. Longley's combinatory algebra of realizers has the 
domain $\GG$ as its carrier, and can be seen as a `coded' version of Berry-Curien 
sequential algorithms \cite{Cur86}; whereas we use the domain of \emph{sequential
  functions on $\GG$} as our carrier, with application defined as in NFGoI, to realize the fully abstract model.

Let us give a first indication of how we can use our simple universal 
(type-free)
space of strategies to recover the same structure we displayed in a typed
version in the previous section.

We split the `address space' by fixing 
\[ C = C_1 \uplus C_2 \]
($C_1$, $C_2$ infinite). Note that
\[ \GG = V_{\bot}^{C} = V_{\bot}^{C_1 + C_2} \cong V_{\bot}^{C_1} \times
V_{\bot}^{C_2} \cong \GG \times \GG . \]
So we can regard any $\sigma \in \Strat$ as
\[ \sigma : \GG \times \GG \longrightarrow \GG \times \GG . \]
The most basic example of a strategy is the \emph{asynchronous copy-cat}:
\[ \mathtt{id}(x, y) = (x \vee y, x \vee y) . \]
The corresponding output function is the twist map:
\[ \mathtt{tw}(x, y) = (y, x) \]
since 
\[ (x \vee y, x \vee y) = (x, y) \vee (y, x) . \]

\subsection{The process model}
This may still be too abstract for some. We can also describe the model in operational
terms with a little process calculus.

We define a syntax of terms by the BNF
\[ P \;\; ::=  \;\; c?x \rightarrow P \mid c!e \mid P \| Q \mid 0 
\mid \top \mid (\nu c)P \mid X \mid \mathtt{rec} \, X. \, P \mid \ldots \]
We define a structural congruence on these terms in the style of the 
(asynchronous) $\pi$-calculus \cite{Mil99}. 
The main novel feature is that $\top$ acts as a zero,
and we have
\[ c!v_1 \, \| \, c!v_2 \equiv \top \;\; (v_1 \neq v_2 ) . \]
We can then define a reduction semantics, again in the style of the
$\pi$-calculus, but with one key difference: we have the reduction
\[ c!v \, \| \, (c?x \rightarrow P)  \;\; \longrightarrow \;\; c!v \, \| \, 
P[v/x] \]
where in CCS or the $\pi$-calculus we would have the rule
\[ c!v \, \| \, (c?x \rightarrow P) \;\; \longrightarrow \;\; P[v/x] . \]
This property of \emph{persistence} of outputs (which corresponds to the
information increasingness in the domain-theoretic presentation) is more
akin to concurrent constraint programming \cite{Sar93}.

We can connect this operational view very directly to our view of strategies
as closure operators by defining a denotational semantics for terms
of our process calculus as closure operators, and proving a correspondence
between this denotational semantics and the operational semantics. We can
then see the process calculus terms as providing a syntax for our
semantic objects, and moreover making the
computational behaviour of strategies explicit via the associated 
operational semantics.

We give a couple of clauses for this denotational semantics.
\[ \lsem c?x \rightarrow P \rsem (a) = \left\{ \begin{array}{ll}
a, & a(c){\uparrow} \\
\lsem P[v/x] \rsem (a), & a(c) = v
\end{array} \right. \]

\[ \lsem c!v \rsem (a) = \left\{ \begin{array}{ll}
a \cup \{ (c, v)\} & \mbox{if consistent} \\
\top & \mbox{otherwise}
\end{array} \right. \]

Parallel composition is interpreted as the join of closure operators;
note the connection to the composition operators defined in Section~5.
\[ \lsem P \| Q \rsem = \bigvee_{k \in \omega} (\lsem P \rsem \circ 
\lsem Q \rsem )^k = \bigvee_{k \in \omega} (\lsem Q \rsem \circ \lsem P \rsem
)^{k} . \] 
Applied to any $a$, this gives the least common fixpoint of $P$ and $Q$
above $a$. Note that it doesn't matter who starts!

Using this calculus, we can construct the copy-cat strategy described
previously as an abstract closure operator in more concrete terms
as a process. (We use the
splitting of the address space introduced in the previous section.
For each cell $c$, we think
of it occurring once in the `left part' of the address space, $C_1$, as
$\mathtt{l}(c)$, and once in the right part $C_2$, as $\mathtt{r}(c)$).
Now we can define the `copy-cat process' as:
\[ 
\mathtt{id} \; \eqdef \;  \|_{c \in C} \; (\mathtt{l}(c)?x \rightarrow \mathtt{r}(c)!x 
\; \| \;
\mathtt{r}(c)?x \rightarrow \mathtt{l}(c)!x) . \]
We can think of this process as a big team of players, one for each 
matched pair of cells.
This process copies the contents of either element of the pair into the other as
soon as it is filled. Of course, if the pair of cells have incompatible
contents, then this will fail. However, if we play this copy-cat strategy 
against any other strategy
following the type constraints of Linear Logic, then this kind of failure
can never happen.

\section{Full Completeness}
The usual notion of completeness for a logic
is with respect to provability; Full Completeness is with respect to
\emph{proofs}.

Let $\MM$ be a model of the formulas \emph{and} proofs of a logic $\LL$.
Typically this means that $\MM$ is a category with structure of an appropriate
kind, such that the formulas of $\LL$ denote objects of $\MM$, proofs
$\Pi$ in $\LL$ of entailments $A \vdash B$ denote morphisms
$$ \lsem \Pi \rsem : \lsem A \rsem \longrightarrow \lsem B \rsem , $$
and the convertibility of proofs in $\LL$ with respect to cut-elimination
is soundly modelled by the equations between morphisms holding in $\MM$.
We say that $\MM$ is \emph{fully complete} for $\LL$ if for all formulas
$A$, $B$ of $\LL$, every morphism $f : \lsem A \rsem \rightarrow \lsem B \rsem$
in $\MM$ is the denotation of some proof $\Pi$ of $A \vdash B$ in $\LL$:
$f = \lsem \Pi \rsem$. Thus the full completeness of $\MM$ means that it
characterizes `what it is to be a proof in $\LL$' in a very strong sense.
If $\MM$ is defined in a syntax-independent way, this is a true semantic
characterization of the `space' of proofs spanned by $\LL$.

The notion of Full Completeness was introduced in
\cite{gfc}\footnote{There were a number of significant precursors, as 
  noted in \cite{gfc}, including representation theorems in category
  theory \cite{FS}, full abstraction results in programming language
  semantics \cite{Mil75,Plo77},
Plotkin's characterization of definability in the $\lambda$-calculus
using logical relations \cite{Plo73},
  studies of parametric polymorphism \cite{BFSS,HRR}, and the completeness conjecture in 
  \cite{Gir91}. However, the contribution of \cite{gfc} was to clearly
  identify and formulate this issue as a precise and interesting
  research programme, and to prove the first in what has become quite
  a rich sequence of results.}, and a Full
Completeness theorem was proved for a game semantics of Multiplicative
Linear Logic (with the MIX rule). This was followed by a series of papers
which established full completeness results 
for a variety of models with respect
to various versions of Multiplicative Linear Logic, e.g.
\cite{HO93,BS,Loa94,ralph-phd,Tan,DHPP,Hag00a}. 

The proofs of full completeness which have appeared to date fall into
two broad classes:
\begin{itemize}
\item Proofs using decomposition arguments
\item `Geometric' proofs via proof nets.
\end{itemize}
The first class of proofs is based on the `last rule = first move'
idea. One decomposes a (sequential) strategy into an opening move, or
an initial P-response to an opening O-move, 
and some sub-strategies. The initial protocol is matched with a
sequent calculus rule, or a term formation scheme in a
$\lambda$-calculus setting. One then recursively applies the
decomposition to the sub-strategies, uncovering the sequent proof or
term which denotes the strategy step by step.

This is typical of the proofs of definability and full completeness
which have been given for a range of $\lambda$-calculus based
programming languages \cite{AMcC99,Abr00}, and for 
various fragments of Intuitionistic Logic and Intuitionistic Linear
Logic \cite{MO99,MO00}, and also applies, \textit{modo grosso}, to the full completeness proof for
Ludics with respect to $\Mallf$ \cite{Gir01}. This form of proof is
obviously well-adapted to sequential models.

By contrast, the second class of proof, beginning with the original
full completeness result in \cite{gfc}, first establishes a connection
beteen the semantic objects (in our case, strategies) and proof
structures, or `pre-proofs'; and then shows that the `geometric'
constraints picking out those structures which are proof nets, i.e
which correspond to real proofs, must be satisfied by the proof
structures arising from the semantic objects.

The first step typically uses uniformity arguments of a fairly general 
nature, while the second step makes more delicate use of specific
features of the model.

The full completeness proof for concurrent games with respect to MALL
\cite{draft,AM99} is of this general form. The connection made is with the proof
nets for MALL introduced in \cite{girard:pn}, which eliminate additive boxes in 
favour of boolean weights, which distribute information about causal
dependencies around the proof net. An important point is that what
seemed to be mere technical machinery in \cite{girard:pn} acquires a much
clearer semantic status in the full completeness argument. In
particular, the boolean weights are derived from inherent functional (or, in
process terms: causal) dependencies in the model; and the `monomial
condition', which plays a key technical r\^{o}le in the proof of the
Sequentialization Theorem in \cite{girard:pn}, turns out to correspond exactly
to the semantic constraint of \emph{stability} on the strategies in the
concurrent games model.

We thus have the interesting situation of two rather different proofs: 
one for the sequential case (Ludics with respect to $\Mallf$) and one
for the concurrent case (concurrent games with respect to
MALL). Nevertheless, the scope of the two results is entirely
analogous: both concern the $\Pi^1$ fragment of MALL resp. $\Mallf$,
without the multiplicative units. It is a challenge for further work
to relate them.

\section{Concluding Remarks}

We have sketched two ways out of the structural dilemma which arose
with Blass games. There are good arguments for pursuing both:

\begin{itemize}
\item The focussing version of Linear Logic offers more control over
  proof search (the purpose for which it was originally introduced
  \cite{AP91}), and is a good setting for studying sequential models of
  classical systems, analogous to the $\lambda \mu$-calculus for
  Classical Logic.

\item The concurrent approach allows the direct study of the original
  system. It is close to Geometry of Interaction and Proof nets. It
  may offer a principled way of describing and controlling true
  concurrency, just as classical logic may for non-deterministic
  computation.
\end{itemize}
Ultimately, one would hope to relate these two directions in a
coherent fashion.

We end with some more specific challenges and questions:

\begin{itemize}
\item Both for Ludics and concurrent games, the challenge of extending 
  the full completeness results to the exponentials remains. This may
  require a better treatment of the multiplicative units than either
  currently offers.

\item Once the exponentials have been analyzed, one may hope to turn
  the lens of fully complete models onto systems which can be
  interpreted into Linear Logic, notably Classical logic. One
  tantalizing question which remains unclear: is there true
  concurrency lurking somewhere in the computational interpretation of 
  classical logic?

\item It would also be interesting to look at bounded
  systems such as Light Linear Logic \cite{LLL}, and the connections to
  complexity classes.
\end{itemize}
\section*{Appendix: a brief review of MALL}

The formulas of the system are built from \emph{literals},
i.e. propositional atoms 
$\alpha$, $\beta$, \ldots and their negations $\alpha^{\perp}$, $\beta^{\perp}$,
\ldots , by the grammar
\[ A \;\; ::= \;\; \alpha \mid \alpha^{\perp} \mid A \otimes A \mid A 
\llpar A \mid A \oplus A \mid A
\& A . \]
Here $\otimes$ and $\llpar$ are the \emph{multiplicative} connectives, 
while $\oplus$, $\&$ are the \emph{additive} connectives.

\noindent Negation is definitionally extended to general formulas by the
equations
\[ (A \otimes B)^{\bot} = A^{\bot} \llpar B^{\bot} \qquad (A \llpar
B)^{\bot} = A^{\bot} \otimes B^{\bot} \]
\[ (A \oplus B)^{\bot} = A^{\bot} \& B^{\bot} \qquad (A \&
B)^{\bot} = A^{\bot} \oplus B^{\bot} \]
\[ A^{\bot \bot} = A . \]
A sequent in MALL is an expression $\girard \Gamma$, where $\Gamma$ is 
a finite multiset of
formulas.

\noindent The rules of MALL are as follows.
\\
\noindent \textbf{Axiom/Cut}
\[ \infer[\Ax]{\girard A, A^{\bot}}{}
\qquad
\infer[\Cut]{\girard \Gamma , \Delta}{\girard \Gamma , A & \girard
  \Delta , A^{\bot}}
\]
\textbf{Multiplicatives}
\[ \infer[\Ten]{\girard \Gamma ,  \Delta , A \otimes B}
{\girard \Gamma , A & \girard \Delta , B}
\qquad
\infer[\LLPar]{\girard \Gamma , A \llpar B}{\Gamma , A, B}
\]
\textbf{Additives}
\[ \infer[\Plusl]{\girard \Gamma  A \oplus B}{\girard \Gamma , A}
\qquad
\infer[\PlusR]{\girard \Gamma  A \oplus B}{\girard \Gamma , B}
\qquad
\infer[\WITH]{\girard \Gamma , A \& B}
{\girard \Gamma , A & \girard \Gamma , B}
\]

\end{document}